\documentclass[12pt,preprint]{aastex}

\usepackage{amsmath}
\usepackage{color}

\newcommand{\myemail}{t-tanigawa@med.uoeh-u.ac.jp}
\newcommand{\pfrac}[2]{\frac{\partial #1}{\partial #2}}
\newcommand{\ppfrac}[2]{\frac{\partial^2 #1}{\partial {#2}^2}}
\newcommand{\kakkoi}[3]{\left(\frac{{#1}}{{#2}}\right)^{#3}}
\newcommand{\e}[1]{\times 10^{#1}}
\newcommand{\erf}{{\rm erf}}
\newcommand{\erfc}{{\rm erfc}}
\slugcomment{Submitted to ApJ}

\shorttitle{Final Masses of Giant Planets}
\shortauthors{Tanigawa \& Tanaka}

\begin{document}


\title{Final Masses of Giant Planets II: Jupiter Formation in a
Gas-Depleted Disk}


\author{Takayuki Tanigawa}
\affil{School of Medicine, University of Occupational and Environmental
Health, Kita-Kyushu, 807-8555, Japan}
\email{\myemail}

\and

\author{Hidekazu Tanaka}
\affil{Institute of Low Temperature Science, Hokkaido University,
    Sapporo, 060-0819, Japan}



\begin{abstract}
Firstly, we study the final masses of giant planets growing in
protoplanetary disks through capture of disk gas, by employing empirical
formulas for the gas capture rate and a shallow disk gap model, which
are both based on hydrodynamical simulations.  We found that, for
planets less massive than 10 Jupiter masses, their growth rates are
mainly controlled by the gas supply through the global disk accretion,
and the gap opening does not limit the accretion.  The insufficient gas
supply compared with the rapid gas capture causes a depletion of the gas
surface density even at the outside of the gap, which can create a disk
inner hole.  Secondly, our findings are applied to the formation of our
solar system.  For the formation of Jupiter, a very low-mass gas disk
with several Jupiter masses is required at the beginning of its gas
capture because of the non-stopping capture.  Such a low-mass gas disk
with sufficient solid material can be formed through viscous evolution
from an initially $\sim$10AU-sized compact disk.  By the viscous
evolution with a moderate viscosity of $\alpha \sim 10^{-3}$, most of
disk gas accretes onto the sun and a widely spread low-mass gas disk
remains when the solid core of Jupiter starts gas capture at $t \sim
10^7$yrs.  A very low-mass gas disk also provides a plausible path where
type I and II planetary migrations are both suppressed significantly.
In particular, the type II migration of Jupiter-size planets becomes
inefficient because of the additional gas depletion due to the rapid gas
capture by themselves.
\end{abstract}





\section{Introduction}
%
%
A leading hypothesis of giant planet formation is the core instability
model \citep{Mizuno80, BP86, Pollack96, Ikoma00, Hubickyj05}.
In a protoplanetary disk, a solid protoplanet attracts the disk gas and
has a proto atmosphere.
When the core solid is around 10 Earth masses, atmospheric mass becomes
comparable to the core mass, and the atmosphere becomes gravitationally
unstable, which triggers dynamical collapse of the atmosphere to become
a giant planet.
Since the gas accretion of the atmosphere onto the core is in an
unstable and runaway manner, the growth continues as long as gas exists
around the planet.

%
The gas-accretion growth of a giant planet is expected to terminate when
the planet creates a gap, which is a low-density annulus region along
the planet orbit, by its own strong gravity when the planet becomes
massive.
Two well-known gap-opening conditions have been widely used: the thermal
condition and the viscous condition \citep{LP93, IdaLin04, Crida06}.
The thermal condition is a condition that the (specific) gravitational
energy at a distance of the disk scaleheight $\sim GM_{\rm p}/h$ is
larger than the typical thermal energy $\sim c^2$, where $G$ is the
gravitational constant, $M_{\rm p}$ is the planet mass, $h$ is the disk
scale height, and $c$ is the sound speed of disk gas.
The viscous condition is a condition that planetary gravitational torque
exerting on the gas disk is stronger than the viscous torque of the
disk due to Keplerian shear motion \citep{LP86}.
Since the thermal condition usually requires a larger planet mass for
gap opening than that of the viscous condition and the required planet
mass is consistent with Jupiter, the final masses of giant planets have
been thought to be determined by the thermal condition.

%
\citet{TI07} (hereafter TI07) have constructed an analytic model for the
gas accretion rate onto a planet embedded in a disk gap as a function of
the planetary mass, viscosity, scale height, and unperturbed surface
density.  By using this, they systematically studied the long-term
growth and the final masses of gas giant planets.
To calculate the accretion rate, TI07 derived an analytic formula for
surface density distribution in the gap region, where planet
gravitational perturbation is significant.
In addition to the gap formula that considers the balance between the
viscous torque and the planetary gravitational torque
\citep[e.g.,][]{Lubow06}, TI07 also included the gap shallowing effect
by the Rayleigh stable condition that inhibits a too steep radial
gradient of surface density.
The shallowing effect supplies a non-negligible amount of gas into the
gap bottom, which enables the giant planet to keep on growing even after
the gap opening.
At the same time, TI07 also proposed that the gas accretion rate onto
the planet can be limited by the disk viscous accretion rate.
An insufficient gas supply by the disk accretion inevitably limits the
gas accretion rate onto the planet even if the planet is capable of
capturing the ambient gas at a higher rate.
Such a limited accretion rate onto a planet was also used in population
synthesis calculations \citep[e.g.,][]{Mordasini09, Mordasini12a}.
As a result, TI07 gave much larger final masses ($\gtrsim 10$ Jupiter
mass at 5AU) than the traditional prediction with the thermal condition
for the minimum mass solar nebula (MMSN) disk.

%
Recent hydrodynamic simulations have extracted an empirical formula for
surface density at the bottom of gap \citep{DM13, Fung14}.  This formula
indicates that the gap is much shallower than the traditional
prediction, and is even shallower than the analytic estimate given by
TI07, which includes the shallowing effect by the Rayleigh stable
condition.
\citet{Kanagawa15a} analytically derived this shallow gap formula, by
including the effect of density wave propagation at the gap.
Such a shallow gap model maintains high accretion rates onto planets and
gives much larger final masses of giant planets than the prediction by
TI07, leading to a possibility that Jupiter and Saturn formed in a much
lighter disk than the MMSN model.

%
The rapid accretion onto a planet due to a shallow gap also causes a
depletion of the disk gas over a wider radial region, in addition to the
narrow gap.  This gas depletion may alter the type II planetary
migration as well as the planet growth.  \citet{Lubow06} examined this
depletion mechanism due to the gas accretion onto a planet, by using
their semi-analytical model and hydro-dynamical simulations.  However,
their estimations might suffer from large errors since the calculation
time of their hydro-dynamical simulations is less than 1/10 of the
characteristic viscous evolution time. It would be valuable to examine
the gas depletion due to the gas accretion onto a planet, by using an
updated formula for the accretion rate with the shallow gap model.

In this study, we update the growth model of giant planets proposed by
TI07, by adopting the empirical shallow gap model, and demonstrate that
the termination of giant planet growth by the gap opening is much harder
than expected in the traditional prediction.
From this result on the growth rate, we propose that a gas depleted disk
is suitable for the formation of Jupiter-sized planets.  We also
estimate the gas depletion due to the rapid gas accretion onto the
planet using the updated formula for the accretion rate, which enables
us to quantitatively discuss the inner hole and its effect on the type
II planetary migration.
We first we describe the formulation of our model in section
\ref{sec:formulation}.  We next shows examples of evolution of gas
capturing growth and final mass of the giant planets in section
\ref{sec:results}.
We discuss a plausible path for formation of Jupiter in section
\ref{sec:implication}.  The type II migration is also discussed there.
Our results are summarized in section \ref{sec:summary}.

\section{Formulation}\label{sec:formulation}
\subsection{Disk model}
We consider a globally evolving protoplanetary disk.
The protoplanetary disk has scaleheight $h=c/\Omega$, where $c$ and
$\Omega$ are the sound speed and the Keplerian angular velocity around
the central star, respectively.
We set temperature distribution so that $h/r = 10^{-1.5} (r/{\rm
1AU})^{1/4}$, where $r$ is the distance from the star.
This corresponds to the temperature profile $T \simeq 280$ K $(r/{\rm
1AU})^{-1/2}$ for solar type stars.
We use $\alpha$-model for disk viscosity: $\nu = \alpha c h$, where
$\alpha$ is a non-dimensional parameter and independent of the radius $r$
and time \citep{SS73}.  For the above temperature profile, $\nu$ is
proportional to $r$.
We adopt a self-similar solution for global evolution of the
protoplanetary disk \citep{Hartmann98, LP74}.  The surface density of
the solution is given by
\begin{equation}
\Sigma_{\rm ss}(r,t)
= \frac{M_{\rm d,ini}}{2\pi R_{\rm o}^2}
  \kakkoi{r}{R_{\rm o}}{-1}
  \tilde{t}_{\rm ss}^{-3/2}
  \exp \left( -\frac{r}{\tilde{t}_{\rm ss} R_{\rm o}}\right),
\label{eq:Sigma_ss}
\end{equation}
\begin{equation}
\tilde{t}_{\rm ss}
 = \frac{t}{\tau_{\rm ss}} + 1,
\label{eq:t_ss}
\end{equation}
where $M_{\rm d,ini}$ is the initial total mass of the protoplanetary
disk and $R_{\rm o}$ is the disk outer radius at $t=0$.  Note that the
initial time, $t=0$, is set to be the onset time for the dynamical gas
accretion onto the giant planet's core in our model.  In Equation
(\ref{eq:t_ss}), we define $\tau_{\rm ss} = R_{\rm o}^2/3\nu_{\rm o}$,
where $\nu_{\rm o}$ is the viscosity at $r=R_{\rm o}$.
The total disk mass of this model is written as
\begin{equation}
M_{\rm d,ss}(t)
= \int_0^\infty
      2\pi r \Sigma_{\rm ss}(r,t) dr
= M_{\rm d,ini} \tilde{t}_{\rm ss}^{-1/2}.
\label{M_d_ss}
\end{equation}
This means that disk mass decreases slowly with time as $\propto
t^{-1/2}$.
To avoid the unrealistic long lasting disk, we introduce an additional
exponential decay for the disk
\begin{equation}
\Sigma_{\rm un}(r,t)
= \Sigma_{\rm ss}(r,t)
  \exp \left( -\frac{t}{\tau_{\rm dep}} \right),
\label{eq:Sigmap}
\end{equation}
and the disk mass is written as a function of time: $M_{\rm d,ini}
\tilde{t}_{\rm ss}^{-1/2} \exp (-t/\tau_{\rm dep})$.  The additional
exponential decay would correspond to some other mechanisms for disk
dissipation, such as photoevaporation by ultraviolet radiation from the
central star or disk wind (see discussion in section \ref{sec:summary}).
We use $\Sigma_{\rm un}$ as the unperturbed disk surface density in this
paper.

The global disk accretion rate of the self-similar solution with the
additional exponential decay at an orbital radius $r$ is given by
\begin{equation}
  \begin{split}
\dot{M}_{\rm d,global}(r,t)
&= \frac{M_{\rm d,ini}}{2\tau_{\rm ss}}
   \left(1 - \frac{r}{\tilde{t}_{\rm ss}R_{\rm o}/2}\right)
   \tilde{t}_{\rm ss}^{-3/2}
   \exp \left( -\frac{r}{\tilde{t}_{\rm ss} R_{\rm o}}\right)
   \exp \left( -\frac{t}{\tau_{\rm dep}}\right)\\
   &= 3\pi \nu \Sigma_{\rm un}(r,t)
   \left(1 - \frac{r}{\tilde{t}_{\rm ss}R_{\rm o}/2}\right),
  \end{split}
\label{eq:mdotdg}
\end{equation}
where the factor $\exp(-t/\tau_{\rm dep})$ is due to the additional
exponential decay.


We put the initial total mass of the protoplanetary disk as
\begin{equation}
M_{\rm d,ini}
= 1.1\e{-1}
  f_{\Sigma,\rm 5AU}
  \kakkoi{R_{\rm o}}{\rm 200AU}{1}
  M_\odot,
\label{eq:Mdini}
\end{equation}
where $f_{\Sigma,\rm 5AU}$ is a parameter and $M_\odot$ is the mass of
the Sun.  When $f_{\Sigma,\rm 5AU}=1$, the initial total disk mass of
Equation (\ref{eq:Mdini}) makes the initial unperturbed surface density
$\Sigma_{\rm un}$(5AU, $t=0$) equal to that of the minimum mass solar
nebula model at 5AU \citep[i.e., $1.7\e{-5}M_\odot/{\rm AU}^2 = 1.4\e{3}$
kg/m$^2$,][]{Hayashi85}.

It has been reported that photoevaporation by far-ultraviolet (FUV)
radiation from a central star can considerably accelerate the dispersal
of the circum-stellar disk \citep[e.g.,][]{Gorti09}. Photoevaporation by
FUV mainly removes the gas at the outer disk with $\gtrsim$100AU and
decreases the disk mass exponentially with time at a relatively early
stage ($\sim 10^6$yr). However, the mass loss rate by this mechanism is
still uncertain at the order-of-magnitude level
\citep[e.g.][]{Alexander14}.  In this paper, thus, we do not include the
effect of photoevaporation by FUV on the disk evolution for simplicity.

\begin{figure}
\epsscale{1.00}
\plotone{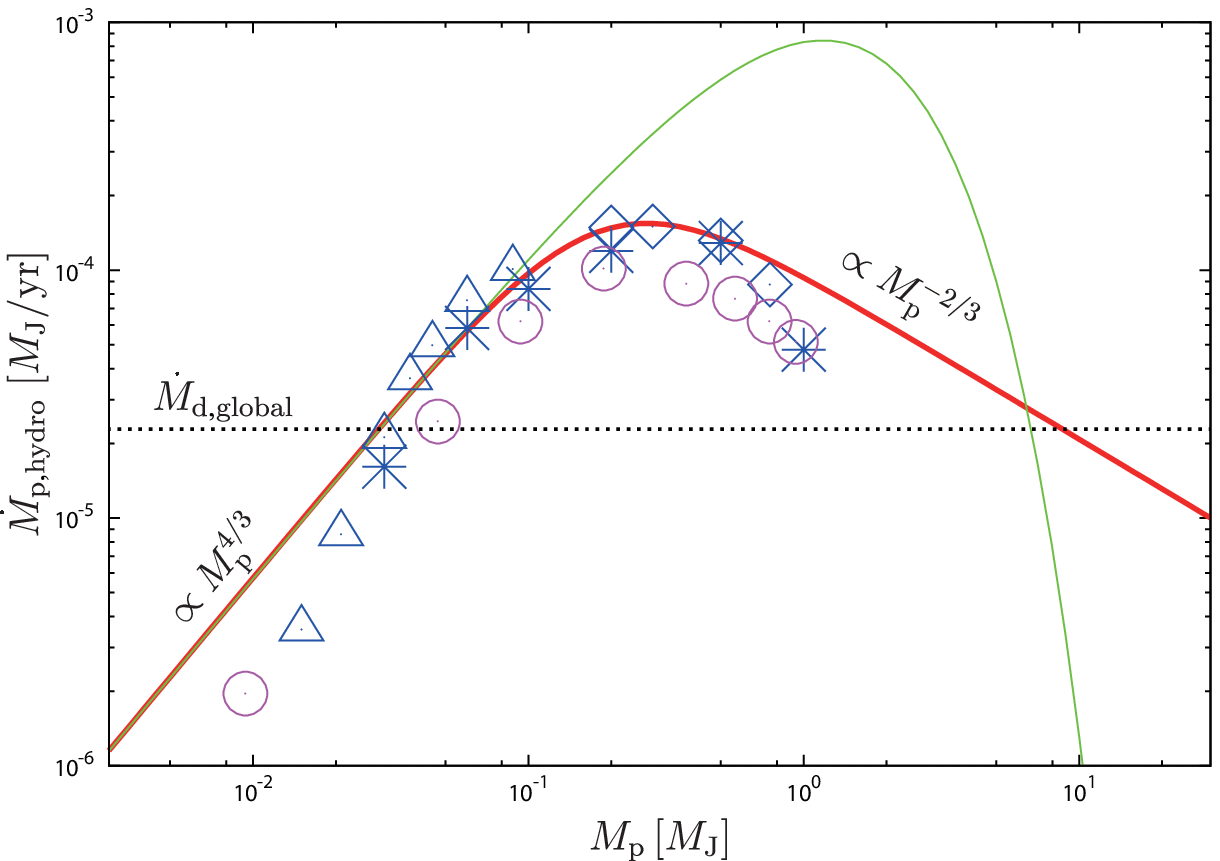}
\caption{
An example of gas accretion rate $\dot{M}_{\rm p,hydro}$ as a function
of planet mass when $\alpha=4\e{-3}$, $h/r_{\rm p}=0.05$, $f_{\Sigma,\rm
5AU}=1$ and $r_{\rm p}=5.2$AU.  The red thick curve is $\dot{M}_{\rm
p,hydro}$ and the thin blue dashed curve is the corresponding accretion
rate derived in \citet{TI07}.  The blue symbols show the accretion rates
obtained by \citet{DAngelo03}, where the three types of the marks
correspond to different models of smoothing planet gravitational
potentials.  The purple circles plot the accretion rates obtained by
\citet{Machida10}.  As a reference, global disk accretion rate
$\dot{M}_{\rm d,global}$ is also shown.
}
\label{fig:mdot_vs_m}
\end{figure}

\subsection{A new simple model for gas accretion onto a planet}
We consider a protoplanet embedded in the evolving protoplanetary disk.
The protoplanet with the mass $M_{\rm p}$ is rotating around the central
star at a distance $r_{\rm p}$ from the star.  The planet starts
dynamical gas capturing, i.e., after gravitational instability of the
proto-atmosphere around a solid core with about 10 Earth masses
\citep{Mizuno80,BP86,Pollack96,Ikoma00,Hubickyj05}.
%

We introduce a formula for gas accretion rate onto the protoplanet from
the protoplanetary disk by an explicit function of parameters, which
enables us to obtain time evolution of the planet mass and eventually
the final mass.  We basically follow the method of TI07, but we have
improved some points, so we will re-summarize it below.

%

%

If sufficient gas is supplied toward the planet orbit by the disk
accretion, the accretion rate onto the gas giant planet is determined by
the hydrodynamics of the gas accretion flow onto the planet and denoted
it by $\dot{M}_{\rm p,hydro}$.  This accretion rate is given by a
product of the two quantities:
%
\begin{equation}
\dot{M}_{\rm p,hydro} = D \Sigma_{\rm acc},
\label{eq:mdotphydro}
\end{equation}
where $D$ is the accretion {\it area} of the protoplanetary disk per
unit time\footnote{$D$ in this paper corresponds to $\dot{A}$ in TI07.},
and $\Sigma_{\rm acc}$ is the surface density at the accretion channel
in the protoplanetary disk.
\citet{TW02} performed two-dimensional hydrodynamic simulations of the
accretion flow onto a planet and derived an empirical formula for the
accretion rate.  According to their result, $D$ is given by
%
\begin{equation}
D
= 0.29
  \kakkoi{h_{\rm p}}{r_{\rm p}}{-2}
  \kakkoi{M_{\rm p}}{M_\ast}{4/3}
  r_{\rm p}^2 \Omega_{\rm p},
\label{eq:Adot}
\end{equation}
where $M_{\rm p}$ and $M_\ast$ are masses of the planet and central
star, respectively, and $h_{\rm p}$ and $\Omega_{\rm p}$ are scaleheight
and Keplerian angular velocity at the planet location, respectively.
Note that the product form of $D$ and $\Sigma_{\rm acc}$ in
Equation~(\ref{eq:mdotphydro}) is valid when equation of state of gas
around the Hill sphere can be approximated by isothermal.

TI07 gave the formula for the surface density $\Sigma_{\rm acc}$ purely
in a theoretical manner, including the Rayleigh stable condition.  This
condition prevents unrealistically too steep surface density gradient
and resultant too deep gap, which is a consequence of the simple
assumption of the balance between viscous torque and gravitational
torque by the planet.
However, recent hydrodynamic simulations showed that the gap is even
shallower than the prediction by TI07 \citep{DM13, Fung14}, which are
also supported by theoretical considerations \citep{Fung14, Kanagawa15a,
Kanagawa15b}.
In this study, we thus use an empirical formula for the gas surface
density at the gap bottom obtained by these studies:
\begin{equation}
\Sigma_{\rm acc}(t)
 = \frac{1}{1+0.034K} \Sigma_{\rm un}(r_{\rm p},t),
\label{eq:Sigma_acc_DM13}
\end{equation}
where
\begin{equation}
K 
= \kakkoi{h_{\rm p}}{r_{\rm p}}{-5}
  \kakkoi{M_{\rm p}}{M_\ast}{2}
  \alpha^{-1}.
\label{eq:KanagawaK}
\end{equation}
In Equation (\ref{eq:Sigma_acc_DM13}), we assumed that the accretion
band is located within the gap bottom.  If the accretion band is located
at the gap edge with a higher surface density, the accretion rate given
by Equations (\ref{eq:mdotphydro}) and (\ref{eq:Sigma_acc_DM13}) would be
an underestimate.
%
%
%
%
We set $M_{\rm p,ini}=3.24\e{-5} M_\ast$ for initial mass of the
protoplanet and $R_{\rm o} = 200$AU.

Figure \ref{fig:mdot_vs_m} shows the accretion rate $\dot{M}_{\rm
p,hydro}$ as a function of the planet mass.  In the low-mass planet case
where the parameter $K$ is much less than 1/0.034 $(\sim 30)$, there is
no gas depletion due to the gap, so $\Sigma_{\rm acc}$ can be simply
replaced by $\Sigma_{\rm un}$ (see Equation~(\ref{eq:Sigma_acc_DM13})).
The accretion rate in this regime is
\begin{align}
 \dot{M}_{\rm p,hydro}
 & =
 \dot{M}_{\rm p,nogap} \nonumber \\
 & \equiv 0.29
  \kakkoi{h_{\rm p}}{r_{\rm p}}{-2}
  \kakkoi{M_{\rm p}}{M_\ast}{4/3}
  \Sigma_{\rm un} r_{\rm p}^2 \Omega_{\rm p}
  \qquad
  \mbox{\rm for $K \ll 1/0.034$}.
\label{eq:mdotnogap}
\end{align}
In the high-mass case where $K \gg 1/0.034$, on the other hand,
$\Sigma_{\rm acc}$ is reduced to $\Sigma_{\rm un}/0.034K$ due to the gap
opening and the accretion rate can be written as
\begin{align}
 \dot{M}_{\rm p,hydro}
 & =
 \dot{M}_{\rm p,gap} \nonumber \\
 & \equiv
  8.5
  \kakkoi{h_{\rm p}}{r_{\rm p}}{1}
  \kakkoi{M_{\rm p}}{M_\ast}{-2/3}
  \Sigma_{\rm un} \nu_{\rm p}
      \qquad
      \mbox{\rm for $K \gg 1/0.034$}.
\label{eq:mdotgap}
\end{align}
Equation (\ref{eq:mdotgap}) shows that the accretion rate $\dot{M}_{\rm
p,hydro}$ decreases gradually $(\propto M_{\rm p}^{-2/3})$ after the gap
opening.

In Figure \ref{fig:mdot_vs_m}, we also plotted the accretion rates
obtained by the previous hydrodynamical simulations to check the validity
of our simple model.
\citet{DAngelo03} examined the gas accretion rate onto a planet embedded
in a protoplanetary disk, by performing three-dimensional global
hydrodynamic simulations for various planet masses.  We find that our
model reproduces well their results.
The results of three-dimensional local simulations by \citep{Machida10}
are also plotted and their results are in good agreement with our
model\footnote{ \citet{Machida10} claimed that the accretion rates in
two-dimensional simulations (TW02), which our model is based on in this
paper, are typically two orders of magnitude larger than those in
three-dimensional cases.  But their fitting formula (Equation~(11) for
$\tilde{r}_{\rm H}^3<0.3$ of their paper) is actually only a factor of 2
(or less) smaller than that in TW02 (Equation~(18) of their paper).
This can be confirmed by the fact that the width and position of the
accretion bands in the two-dimensional (Fig.~8 of TW02) and
three-dimensional \citep[Fig.~3 of][]{TOM12} cases are almost the same
except near the midplane.}.
The accretion rate used in TI07 are also plotted.  TI07's accretion rate
declines rapidly with increasing mass because of its deeper gap model.
The global disk accretion rate $\dot{M}_{\rm d,global}$ is also shown as
a reference.

We also consider the case where the gas supply by the viscous disk
accretion is insufficient.  In such a case, the gas accretion onto the
planet is regulated by the global disk accretion rate $\dot{M}_{\rm
d,global}(r_{\rm p})$ rather than $\dot{M}_{\rm p,hydro}$.  We need to
take into account this effect since the gap opening cannot significantly
slow down the gas accretion onto the planet.  Furthermore, at an early
stage of the gas capture by the planet, an additional treatment is
required for the realistic gas supply to the planet orbit.  At the early
stage, a substantial amount of gas still exists near the planet orbit.
The gas supply from the nearby part is regulated by a local disk
diffusion rather than the global disk accretion.  The disk accretion
rate due to the local diffusion is given by
\begin{equation}
\dot{M}_{\rm d,local}
 = \pi r_{\rm p} \Sigma_{\rm un}(r_{\rm p})
   \sqrt{\frac{\nu_{\rm p}}{t-t_{\rm gap}}},
\label{eq:mdotdl}
\end{equation}
(see Appendix A)\footnote{The coefficient of Equation~(\ref{eq:mdotdl})
is a factor of two smaller than that in Appendix A.  However this factor
does not affect the final results because the accretion rate in Phase 2
(see Section~\ref{sec:results}) is not important for the final mass.}.
In practice, the gas supply would be approximately given by the larger
one of $\dot{M}_{\rm d,global}$ and $\dot{M}_{\rm d,local}$.  In our
model, therefore, by including the gas supply to the planet orbit, we
give the gas accretion rate onto the planet, $\dot{M}_{\rm p}$, as
\begin{equation}
 \dot{M}_{\rm p}
  = {\rm min}( \dot{M}_{\rm p,hydro},
              {\rm max}(\dot{M}_{\rm d,global}, \dot{M}_{\rm d,local})).
\label{eq:mdotp}
\end{equation}

%
%

Using this model for the gas accretion rate onto the planet, we can
easily simulate evolution of the planet mass (or gas accretion rate) for
a given set of disk parameters.  To do that, we only need to numerically
integrate the ordinary differential equation because the integrand is an
explicit function of the disk parameters.  The final mass of a planet is
simply obtained by
\begin{equation}
M_{\rm p,final}(r_{\rm p}, \alpha, h_{\rm p}, M_{\rm d,ini}, R_{\rm o})
 = \int_0^\infty
       \dot{M}_{\rm p}(M_{\rm p}, r_{\rm p}, \alpha, h_{\rm p}, M_{\rm d,ini},
       R_{\rm o}, t) dt.
\label{eq:M_p_final}
\end{equation}

Note that \citet{Fung14} derives a more elaborate fitting formula by
two-dimensional hydrodynamic simulations, which focus on cases for
planets more massive than that of \citet{DM13}.  But the difference
between the two formula is much smaller than that between \cite{DM13}
and TI07, so we use the above equation for simplicity.
In our model, we do not consider radial migration of planets, which will
be discussed in Section \ref{sec:migration}.


\begin{figure}
\epsscale{1.00}
\plotone{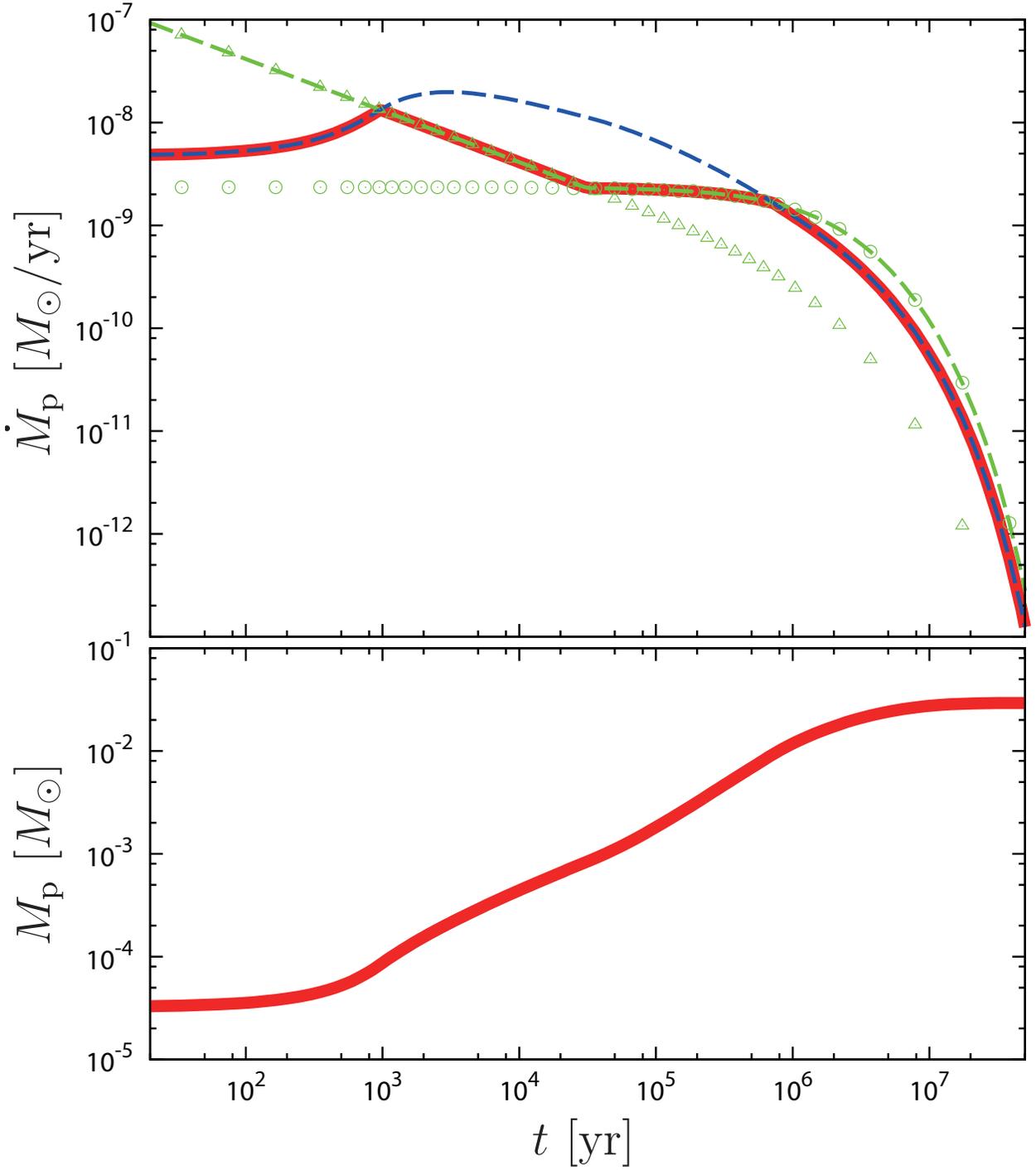}
\caption{An example of time evolution for a protoplanet after the onset
 of dynamical gas capture in the case with $\alpha=3.2\e{-3}$,
 $f_{\Sigma,\rm 5AU}=1$, $\tau_{\rm dep} = 10^7$yr, $R_{\rm o}=200$AU,
 $r_{\rm p}=5$AU.  Top panel shows the gas accretion rate onto the
 protoplanet.  The thick red line shows that the accretion rate that we
 adopt (Equation (\ref{eq:mdotp})), which is a smaller one of
 $\dot{M}_{\rm p,hydro}$ (blue dashed line) or ${\rm max}(\dot{M}_{\rm
 d,global}, \dot{M}_{\rm d,local})$ (green dashed line).  Open circles
 show $\dot{M}_{\rm d,global}$ and open triangles show $\dot{M}_{\rm
 d,local}$.  Bottom panel shows the mass of the protoplanet.}
\label{fig:mdot_vs_t_large_Mp}
\end{figure}

\begin{figure}
\epsscale{1.00}
\plotone{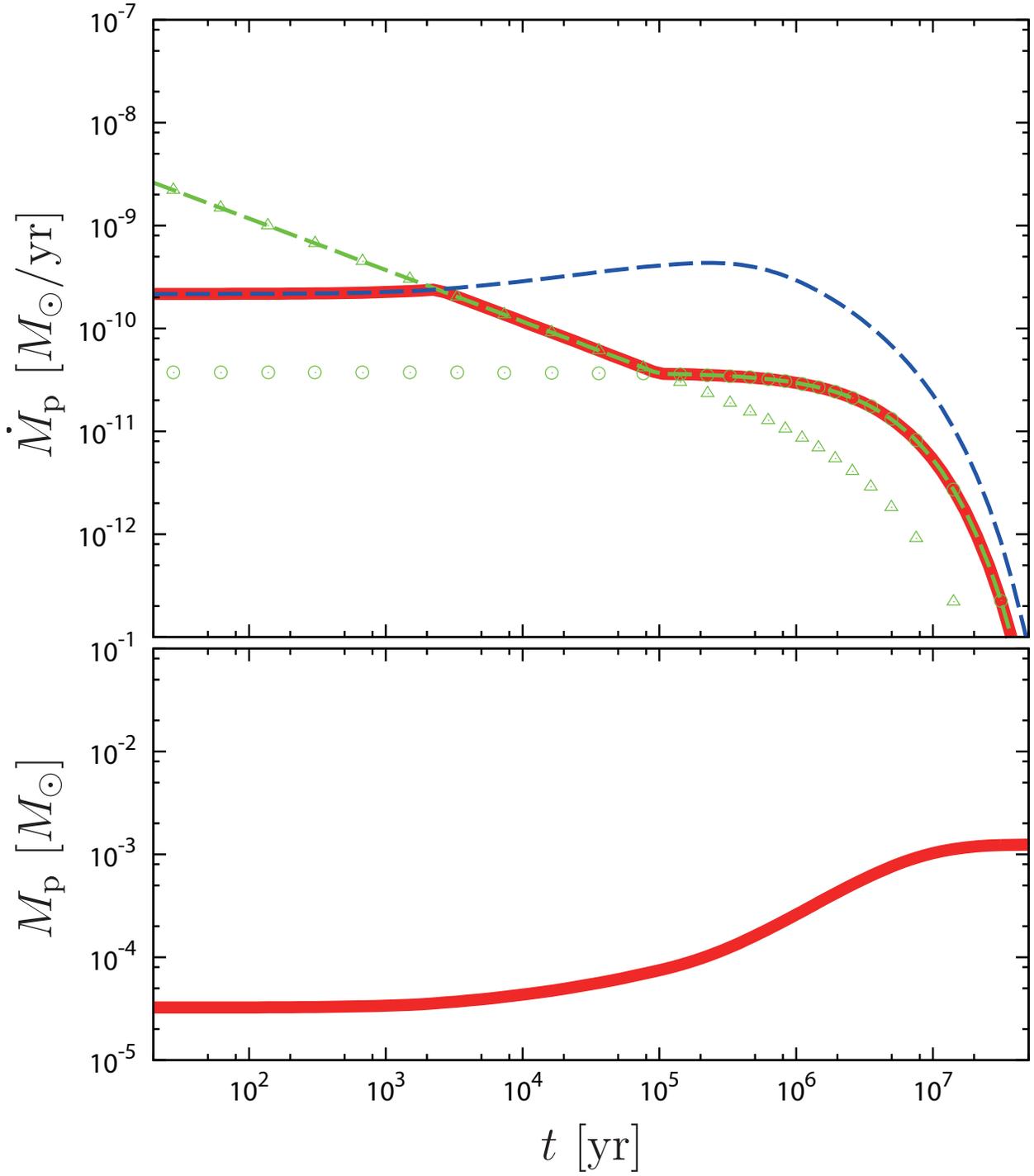}
 \caption{Same as Figure \ref{fig:mdot_vs_t_large_Mp} but in the case
 with $\alpha=10^{-3}$, $f_{\Sigma,\rm 5AU}=1/20$, $\tau_{\rm dep} =
 10^7$yr, $R_{\rm o}=200$AU, $r_{\rm p}=5$AU.}
\label{fig:mdot_vs_t_Jovian_Mp}
\end{figure}

\clearpage
\section{Results}\label{sec:results}
\subsection{Examples of time evolution}\label{sec:evolution}
Figure \ref{fig:mdot_vs_t_large_Mp} plots an example for time evolution
of the gas accretion rate onto a proto-giant planet located at 5AU and
that of planet mass.
The parameters are set to be $\alpha=3.2\e{-3}$, $f_{\Sigma,\rm 5AU}=1$,
$\tau_{\rm dep} = 10^7$yr, $R_{\rm o}=200$AU, $r_{\rm p}=5$AU.
This example illustrates that the evolution can be divided into four
phases:
\begin{itemize}
 \item Phase 1: after the onset of dynamical gas accretion, the gas
       accretion rate is regulated by the hydrodynamic accretion flow
       without a gap: $\dot{M}_{\rm p,nogap}$.  In this case, there is
       abundant gas near the planet, and nothing to limit the accretion
       flow.

 \item Phase 2: the gas supply from the nearby part of the planet orbit,
       $\dot{M}_{\rm d,local}$, limits the accretion rate onto the
       planet because it is lower than $\dot{M}_{\rm p,hydro}$ and
       higher than the global disk accretion rate $\dot{M}_{\rm d,global}$.

 \item Phase 3: this is also the case where the gas supply toward the
       planet orbit limits the accretion rate, but the rate is given by
       $\dot{M}_{\rm d,global}$.

 \item Phase 4: this is again the case when gas supply is regulated by
       the hydrodynamic accretion flow, but with a deep gap:
       $\dot{M}_{\rm p,gap}$.
\end{itemize}
In this case, the final mass of the giant planet is as massive as
$30M_{\rm J}$, which is about 1/3 of initial disk mass (see Equation
(\ref{eq:Mdini})).  The parameter set in this case is not very special
(neither heavy nor highly viscous), but still results in forming a
massive planet.  This is because a gap does not significantly suppress
gas accretion onto the planet.

The next example shown in Figure \ref{fig:mdot_vs_t_Jovian_Mp} is a case
that produces a Jupiter-size planet.  This case adopts a lower viscosity
of $\alpha=10^{-3}$ and a much lighter disk mass of $f_{\Sigma,\rm 5AU}
= 1/20$.  Because of the lower surface density, accretion rate in phase
1 is lower and, as a result, accretion rate and planet mass do not
significantly increase until the end of phase 1.
In phases 2 and 3, sequence of the phase transition is basically the
same as the previous example, but the absolute values of accretion rates
are reduced because of the lower surface density and lower viscosity.
Accretion rates in phase 2 and 3 are proportional to $\Sigma
\alpha^{1/2}$ and $\Sigma \alpha$, respectively, thus comparing with the
case of Figure \ref{fig:mdot_vs_t_large_Mp}, the accretion rates in this
case are reduced by factors of 36 and 60, respectively.
As a result of these low accretion rates, mass of the planet does not
significantly increase, and the planet is not able to open a deep gap,
which leads to no emergence of phase 4.
When $r_{\rm p} \ll \tilde{t}_{\rm ss} R_{\rm o}/2$, we obtain from
Equations (\ref{eq:mdotdg}) and (\ref{eq:mdotgap}) as
\begin{equation}
\frac{\dot{M}_{\rm p,gap}}{\dot{M}_{\rm d,global}}
\sim 0.90\times
     \kakkoi{M_{\rm p}}{M_\ast}{-2/3}
     \kakkoi{h_{\rm p}}{r_{\rm p}}{1}.
\label{eq:mdotgap_mdotglobal}
\end{equation}
The emergence of phase 4 requires $\dot{M}_{\rm p,gap} < \dot{M}_{\rm
d,global}$, which gives
\begin{equation}
M_{\rm p}
> 1.0\e{-2} \kakkoi{h_{\rm p}/r_{\rm p}}{0.05}{3/2} M_\ast,
\label{eq:mp_for_phase4}
\end{equation}
which explains the presence or absence of phase 4 in the two cases.
Thus, up to $\sim 10M_{\rm J}$, the reduction of the accretion rate by
the gap is not effective.  Even in phase 4 of Figure
\ref{fig:mdot_vs_t_large_Mp}, $\dot{M}_{\rm p}$ is not so small compared
with $\dot{M}_{\rm d,global}$ because the ratio $\dot{M}_{\rm
p,gap}/\dot{M}_{\rm d,global}$ depends weakly on $M_{\rm p}$ as in
Equation (\ref{eq:mdotgap_mdotglobal}).

\subsection{Final masses}
\begin{figure}
\epsscale{1.00}
\plotone{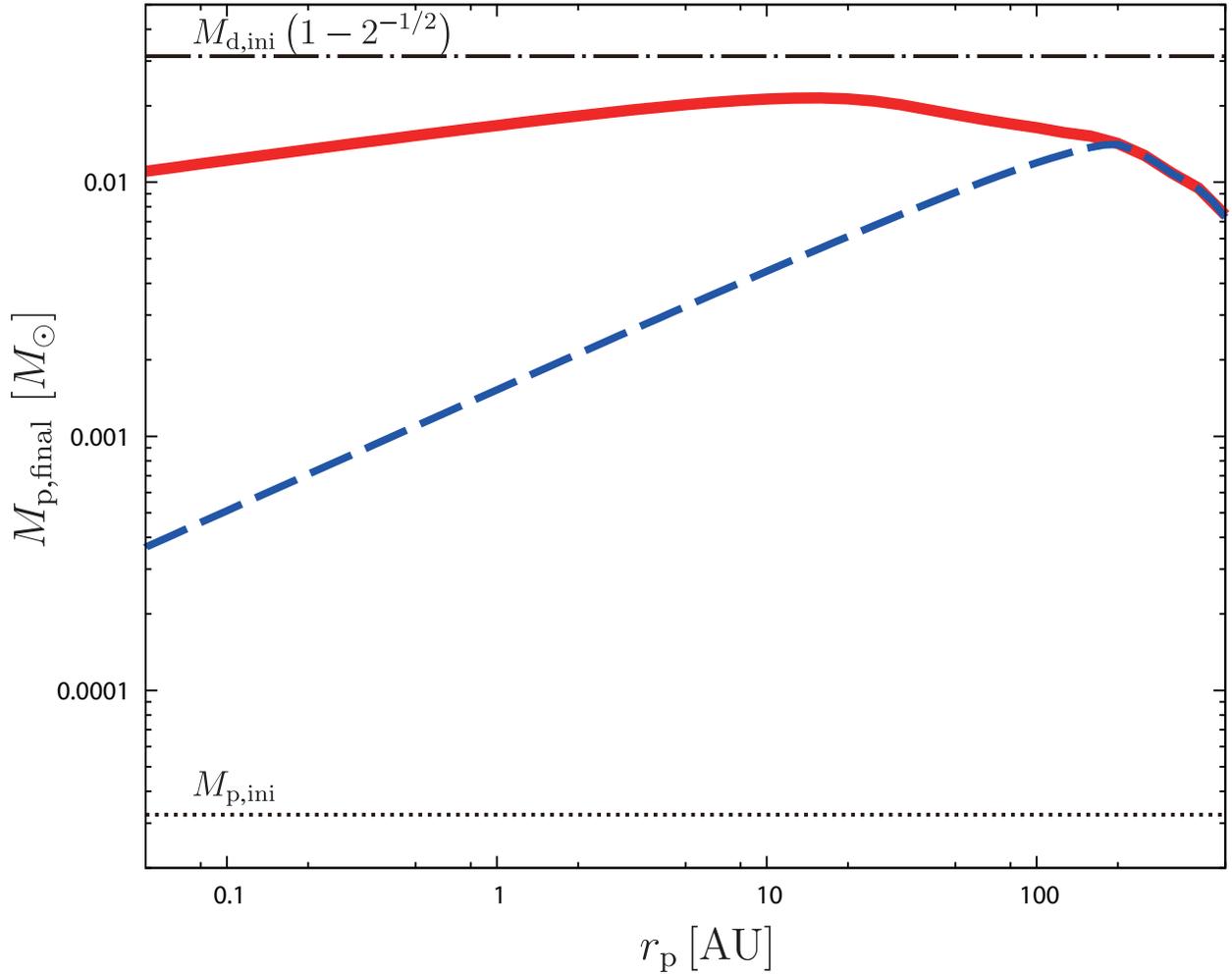}
\caption{Final masses of giant planets as a function of orbital radius
 of the planets when $f_{\Sigma,\rm 5AU}=1$, $\alpha=10^{-3}$, $R_{\rm
 out} = 200$ AU, $\tau_{\rm dep} = 10^7$yr.  Red solid curves adopt the
 model in this paper, and the blue dashed curves shows the case when
 $\Sigma_{\rm acc}$ uses a formula of \citet{TI07}, instead of Equation
 (\ref{eq:Sigma_acc_DM13}).  The dot-dashed line corresponds to the
 final mass that assumes that all the gas accretion is done in phase 3
 (see Equation (\ref{eq:M_p_final_p3})) and $\tau_{\rm dep} = \tau_{\rm
 ss}$, and the dotted line shows the initial mass of the protoplanet
 $M_{\rm p,ini}$.
\label{fig:M_final_vs_r_TI07_vs_DM13}}
\end{figure}

%
Figure \ref{fig:M_final_vs_r_TI07_vs_DM13} plots the final mass of a
gas-capturing planet as a function of orbital radius when $f_{\Sigma,\rm
5AU}=1$, $\alpha=10^{-3}$, $R_{\rm out} = 200$ AU, $\tau_{\rm dep} =
10^7$yr.
The solid line shows the final mass obtained in our model.
We find that the final mass is 10-20 Jupiter masses in most of area and
has only a slight radial dependence.  Up to $\sim 10 M_{\rm J}$, a giant
planet grows mostly in phase 3 and the growth rate is regulated by
global disk accretion rate $\dot{M}_{\rm d,global}$ at all radii.  Thus
the growth rate of giant planets is independent of their radial
location.  Even in phase 4 where $M_{\rm p} \gtrsim 10M_{\rm J}$, the
growth rate is not much smaller than $\dot{M}_{\rm d,global}$.
%
%
We also plot the final mass in the case of TI07 for comparison.  In the
case of TI07, $\Sigma_{\rm acc}$ uses a formula of TI07, while Equation
(\ref{eq:Sigma_acc_DM13}) is used in this paper.
Final mass with TI07's formula becomes larger than Jupiter mass for most
of the region.  This is mainly because TI07 considers the violation of
the Rayleigh condition for steep radial density gradient, which limits
the gradient, tends to fill the gap, and promotes gas capturing growth
as a result.
In this case, the final mass of a planet increases with its orbital
radius $r_{\rm p}$.  This is because gap opening is easier at the inner
region in TI07.  Thus the difference in the final mass between the two
cases is originated from formulae for the gap depth.

The final mass can be estimated with a simple equation.  Since the gap
opening does not significantly affect $\dot{M}_{\rm p}$ up to $\sim
10M_{\rm J}$, the global disk accretion rate $\dot{M}_{\rm d,global}$
determines the final mass in most cases.  Thus, using $\dot{M}_{\rm
d,global}$ for a planet at $r_{\rm p} \ll R_{\rm o}$, final mass is
approximated by
\begin{eqnarray}
 M_{\rm p,final,p3}
&\sim& \int_0^\infty
       \left. \dot{M}_{\rm d,global} \right|_{r_{\rm p}\ll R_{\rm o}}dt
       \nonumber \\
 &\sim&
  M_{\rm d,ini}
  \left[ 1 - \left( \frac{\tau_{\rm dep}}{\tau_{\rm ss}} + 1
	     \right)^{-1/2}
		 \right]
  = M_{\rm d,ini} - M_{\rm d,ss}(\tau_{\rm dep}).
\label{eq:M_p_final_p3}
\end{eqnarray}
This means that all the mass lost from the disk is captured by the
planet.  In the case where $\tau_{\rm dep} \ll \tau_{\rm ss}$, the final
mass of Equation (\ref{eq:M_p_final_p3}) is approximately given by
$(\tau_{\rm dep}/2\tau_{\rm ss}) M_{\rm d,ini}$.
When $r_{\rm p}$ is small, the deviation from
Equation~(\ref{eq:M_p_final_p3}) becomes larger.  This is because the
accretion state is switched from phase 3 to 4 before the end of the
growth.  In this case, the final mass is roughly estimated by
\begin{equation}
 M_{\rm p,final,p4}
  \sim M_{\rm p,final,p3}
  \frac{\dot{M}_{\rm p,gap}(M_{\rm p,final,p3})}
       {\dot{M}_{\rm d,global}}.
\label{eq:M_p_final_p4}
\end{equation}
Note that $M_{\rm d,ini}$ is the disk mass at the time when the planet
starts its gas capture and can be much smaller than the mass when the
disk is formed.  Time scale of global viscous evolution $\tau_{\rm ss}$
in our fiducial case is
\begin{equation}
\tau_{\rm ss}
= \frac{R_{\rm o}^2}{3\nu_{\rm o}}
= 1.1\e{7}
  \kakkoi{\alpha}{10^{-3}}{-1}
  \kakkoi{h_{\rm 1AU}/{\rm 1AU}}{10^{-1.5}}{-2}
  \kakkoi{R_{\rm o}}{\rm 200AU}{1}
  {\rm yr}.
\label{eq:tau_ss}
\end{equation}

\begin{figure}
\epsscale{1.00}
\plotone{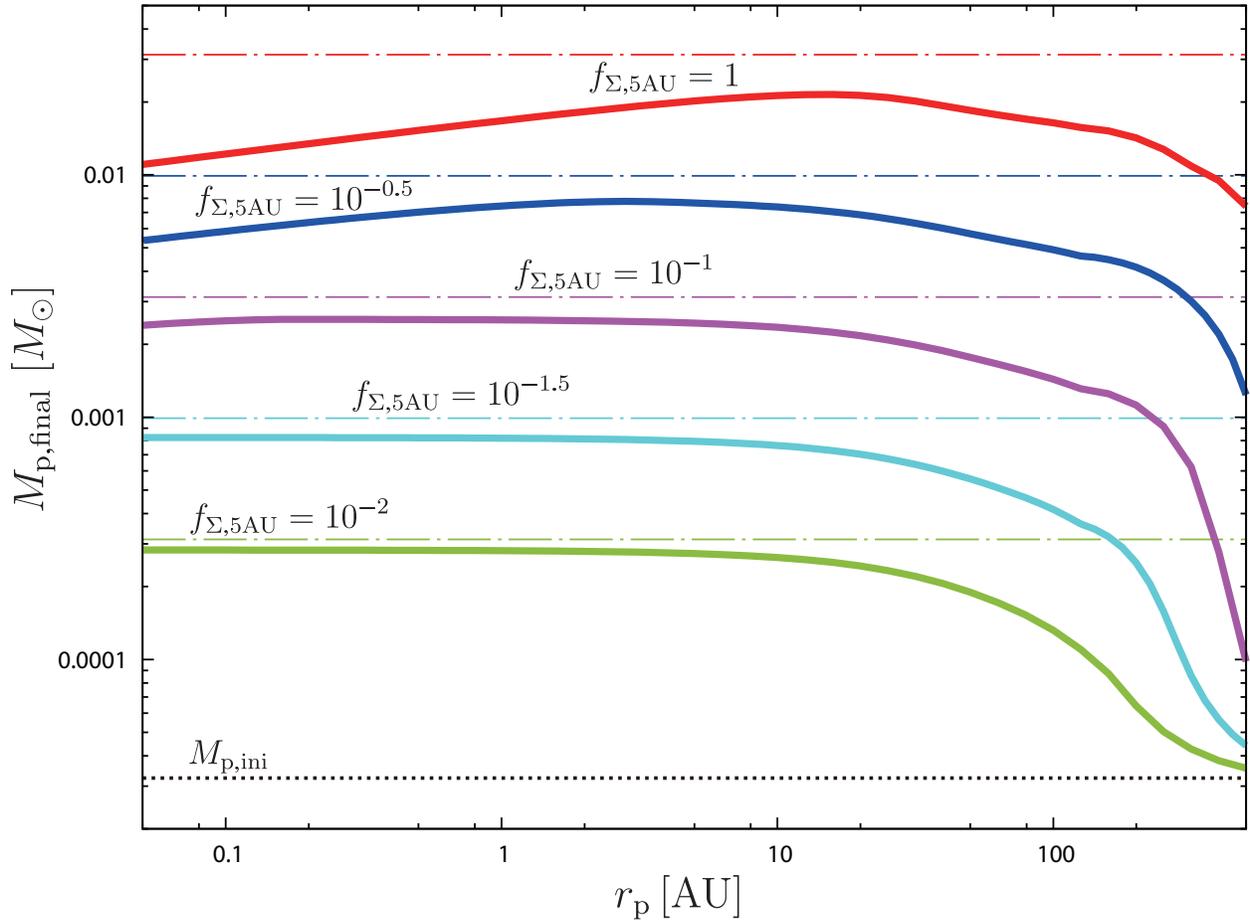}
\caption{Final masses of giant planets as a function of orbital radius
 of the planets when $\alpha=10^{-3}$, $R_{\rm out} = 200$ AU,
 $\tau_{\rm dep} = 10^7$yr.  Five solid curves correspond to
 $f_{\Sigma,\rm 5AU} = 1, 10^{-0.5}, 10^{-1}, 10^{-1.5}, 10^{-2}$ from
 top to bottom, and the dot-dashed line just above the each curve is the
 final mass estimated by Equation (\ref{eq:M_p_final_p3}).
 The dotted line shows $M_{\rm p,ini}$.
\label{fig:M_final_vs_r}}
\end{figure}

Figure \ref{fig:M_final_vs_r} plots final masses for various initial
surface densities (or disk masses).
%
The five curves, which correspond to five $f_{\Sigma,\rm 5AU}$, show
that final mass is proportional to $f_{\Sigma,\rm 5AU}$ in general.
This is simply because the growth rate in phase 3 $\dot{M}_{\rm
d,global}$, which is proportional to surface density, mainly determines
the final mass (see Equation (\ref{eq:M_p_final_p3})), and the gap
effect is not significant.
The final masses shown in this figure are close to possible maximum
masses.  In the case of Figure \ref{fig:M_final_vs_r}, $\tau_{\rm ss}
\simeq \tau_{\rm dep}$, so $M_{\rm d,ini} - M_{\rm d,ss}(\tau_{\rm dep})
= M_{\rm d,ini} (1-1/\sqrt{2})$, which means that all the disk gas
accreting inward is captured by the planet on the way toward the central
star and gap has little effect on suppressing the gas capture.
The final mass in the case of $f_{\Sigma,\rm 5AU}=1$ would be about
$20M_{\rm J}$ around 5AU.  For the formation of Jupiter-mass planets,
the gas disk should be therefore much less massive than the MMSN disk at
the onset of their gas capture.

\begin{figure}
\epsscale{1.00}
\plotone{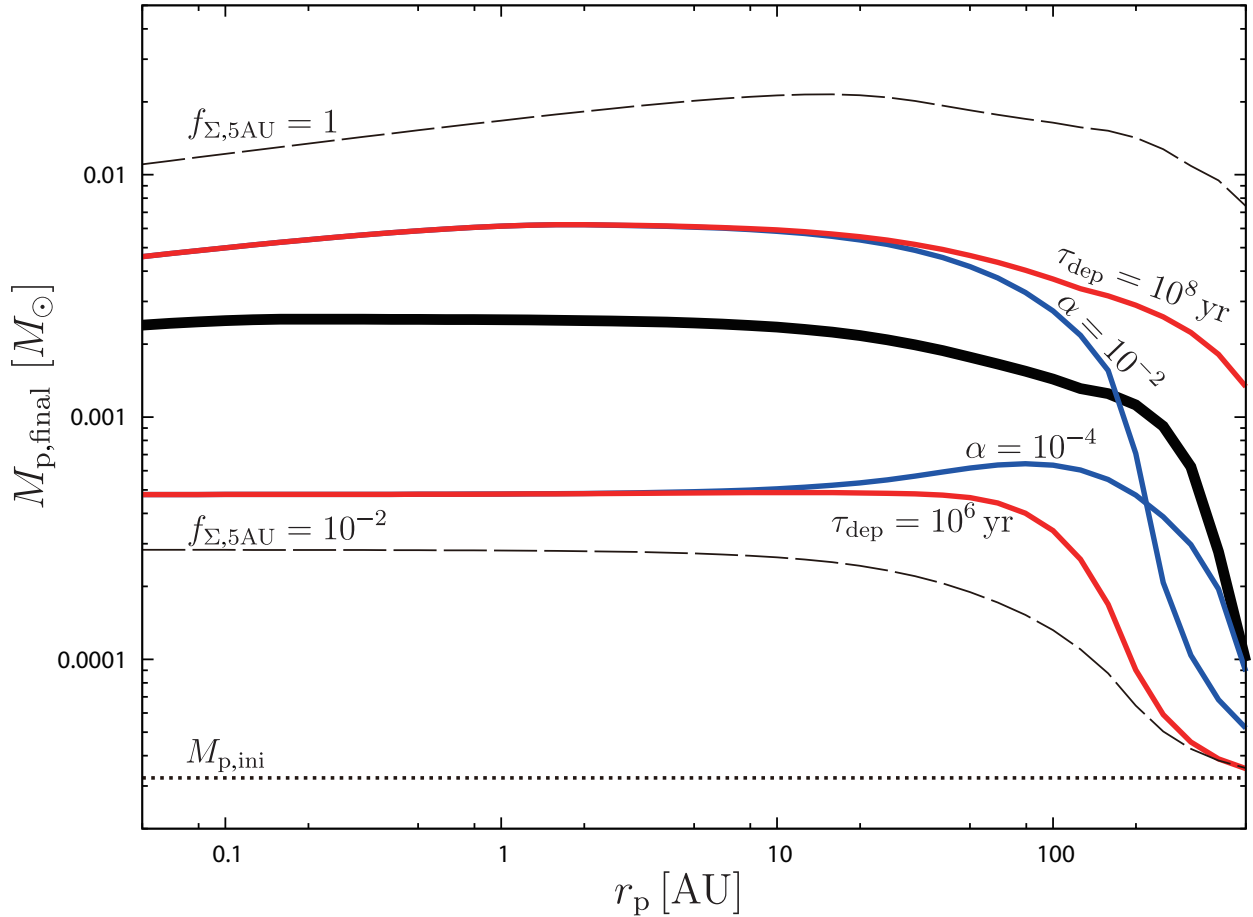}
\caption{Final masses of giant planets as a function of orbital radius
 of the planets.  The thick solid line is the same as the purple line in
 Figure \ref{fig:M_final_vs_r} ($f_{\Sigma,\rm 5AU}=10^{-1}$,
 $\alpha=10^{-3}$, $\tau_{\rm dep}=10^7$yr, $R_{\rm out}=200$AU).  All
 the other lines correspond to cases with 10 times larger (or smaller)
 values for one of the three parameters: $\alpha$, $\tau_{\rm dep}$, and
 $f_{\Sigma, \rm 5AU}$.  The two red lines show $\tau_{\rm dep}=10^8$yr
 and $10^6$yr cases, the two blue lines show $\alpha=10^{-2}$ and
 $10^{-4}$ cases, and the two thin dashed lines show $f_{\Sigma,\rm
 5AU}=1$ and $10^{-2}$ cases.
\label{fig:M_final_vs_r_tau-alpha_depend}}
\end{figure}

Figure \ref{fig:M_final_vs_r_tau-alpha_depend} plots $M_{\rm p,final}$
as a function of $r_{\rm p}$ in the cases with 10 times larger and
smaller values of one of the three parameters: $\alpha$, $\tau_{\rm
dep}$, and $f_{\Sigma, \rm 5AU}$.
We can see that the final mass increases with both $\alpha$ and
$\tau_{\rm dep}$ and depends only on the product $\alpha \tau_{\rm dep}$
in most range.  For example, the degeneracy occurs at $r_{\rm p}
\lesssim 10$AU in the cases with $(\alpha,\tau_{\rm dep}) = (10^{-3},
10^8)$ and $(10^{-2}, 10^7)$ or the cases of $(10^{-3}, 10^6)$ and
$(10^{-4}, 10^7)$.
This is because the final mass is a function of $(\tau_{\rm
dep}/\tau_{\rm ss})$, which is proportional to $\alpha \tau_{\rm dep}$
(see Equation (\ref{eq:M_p_final_p3})).
However, this dependency is weaker than that of $f_{\Sigma,\rm 5AU}$
because the dependence of final mass on $(\tau_{\rm dep}/\tau_{\rm ss})$
is weaker than linear (see Equation (\ref{eq:M_p_final_p3})), while that
on $f_{\Sigma,\rm 5AU}$ is basically linear.
Note that final masses for a pair of the degenerated cases ($\alpha
\tau_{\rm dep} = 10^{-3}$ or $10^{-5}$) are split at $r_{\rm p} \sim
R_{\rm o}$.
This is because most of gas accretion is done by $\dot{M}_{\rm d,local}$
(i.e., phase 2), which is proportional to $\alpha^{1/2}$, not like
$\dot{M}_{\rm d,global} \propto \alpha^1$.  This situation is realized
in the case when $r_{\rm p} \sim R_{\rm o}$ and $\tau_{\rm dep} <
\tau_{\rm ss}$.

\subsection{Gas depletion due to the accretion onto the planet}
In phase 3, the disk gas is further reduced, in addition to the effect
of the gap produced by the planetary torque.  In this phase, the gas
supply by the global disk accretion is insufficient for the rapid gas
capture, which causes an additional depletion of the gas surface density
even at the outside of the narrow gap region.
In this phase, the accretion rate, $D \Sigma_{\rm acc}$, cannot be
larger than $\dot{M}_{\rm d,global}$.  This indicates that $\Sigma_{\rm
acc}$ should be depleted because $D$ is independent of the surface
density (see Equation~(\ref{eq:mdotphydro})).
The additional depletion factor due to the gas capture, $f'$, is
obtained from the balance of the mass fluxes (i.e., the mass
conservation).  Including this depletion factor, the disk surface
density at the outside of the gap is given by $f' \Sigma_{\rm un}(r_{\rm
p})$ and, thus the hydrodynamical capture rate should be evaluated to be
$f' \dot{M}_{\rm p,hydro}$ instead of $\dot{M}_{\rm p,hydro}$.  Assuming
the quasi-steady flow in the disk, we obtain an equation of the mass
flux balance, $f' \dot{M}_{\rm p,hydro} = \dot{M}_{\rm d,global}$.
Hence the additional depletion factor is given by
\begin{equation}
  \begin{split}
   f'
   &= \frac{\dot{M}_{\rm d,global}}
           {\dot{M}_{\rm p,hydro}} \\
   &\simeq 1.1 \kakkoi{M_{\rm p}}{M_\ast}{2/3}
               \kakkoi{h_{\rm p}}{r_{\rm p}}{-1},
  \end{split}
\label{eq:fdash}
\end{equation}
where we used Equation (\ref{eq:mdotgap_mdotglobal}).  From Equation
(\ref{eq:fdash}), the additional depletion factor is unity for $M_{\rm
p}=0.01 M_\ast$ and 0.2 for $M_{\rm p}=M_{\rm J}$ in a disk with $h_{\rm
p}/r_{\rm p}=0.05$.  Note that this additional effect of gas depletion
(or enhance) does not exist in phase 4 because of sufficient supply by
the global disk accretion.

\begin{figure}
\epsscale{1.00}
\plotone{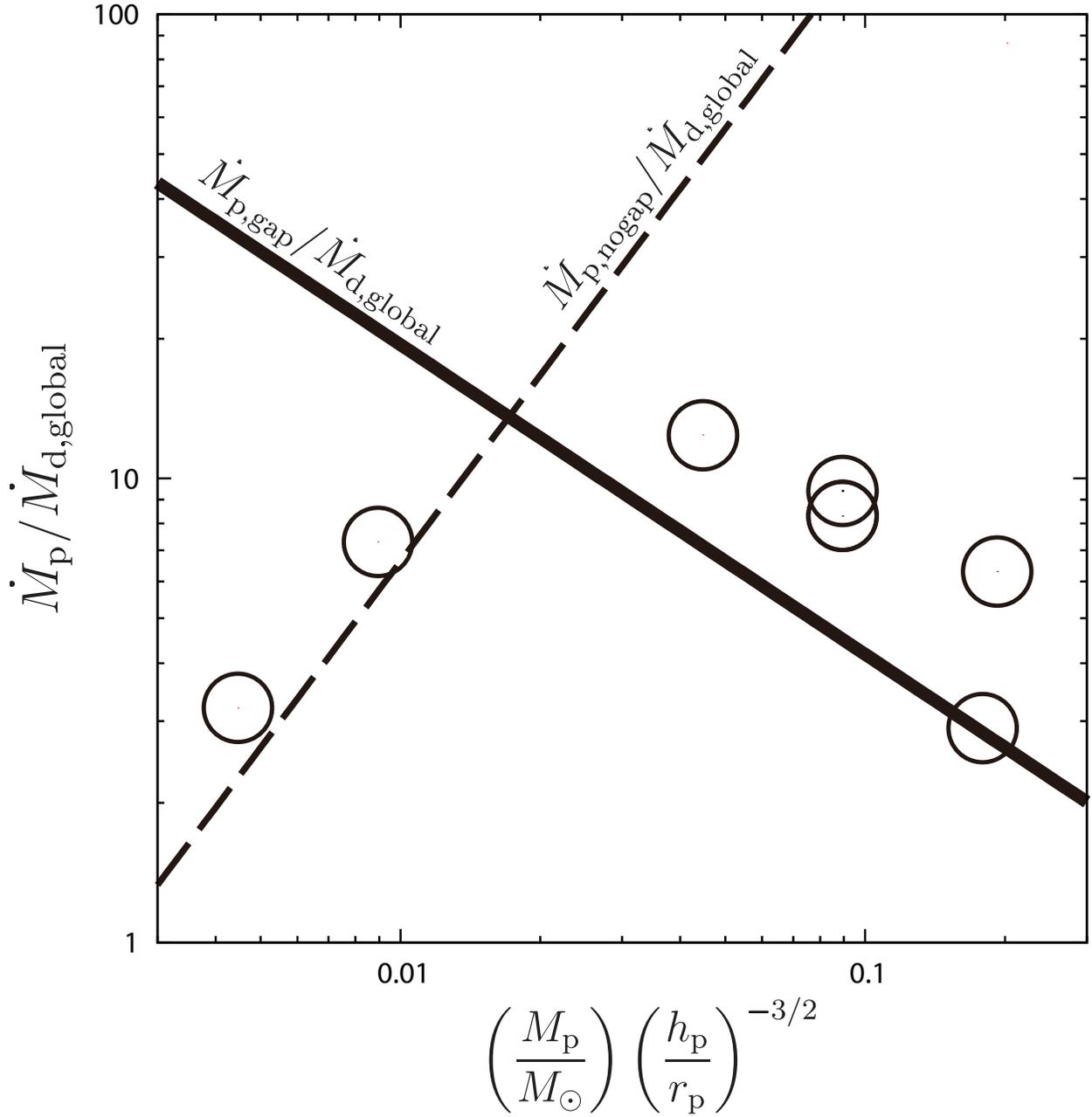}
\caption{Accretion rate onto a planet normalized by disk viscous
 accretion rate.  Circles show accretion efficiency $E$ in Table~1 of
 \citet{Lubow06}, and the solid line shows $\dot{M}_{\rm p,gap} /
 \dot{M}_{\rm d,global}$.  As a reference, $\dot{M}_{\rm p,nogap} /
 \dot{M}_{\rm d,global}$ is also shown by the dashed line.
\label{fig:Lubow_DAngelo_2006_Table1}}
\end{figure}

\citet{Lubow06} also examined the gas depletion due to the accretion
onto the planet and derived the radial distribution of the surface
density, by considering a steady viscous accretion disk with a mass sink
by the planet.
From this accurate surface density distribution, the additional
depletion factor is given by
\begin{equation}
 f' =
  \frac{\dot{M}_{\rm d,global}}
       {(\dot{M}_{\rm p,hydro} + \dot{M}_{\rm d,global})}.
\label{eq:fdash2}
\end{equation}
This agrees with Equation (\ref{eq:fdash}) when $\dot{M}_{\rm p,hydro}
\gg \dot{M}_{\rm d,global}$.
Since a detail derivation was not given in their paper, we presented the
derivation of the surface density distribution in Appendix B.
In \citet{Lubow06}, the ratio
$\dot{M}_{\rm p, hydro} / \dot{M}_{\rm d,global}$ is called the
accretion efficiency. They estimated the accretion efficiency from their
two-dimensional hydrodynamical simulations.  In Figure
\ref{fig:Lubow_DAngelo_2006_Table1}, we plot their results and our model
(i.e., Equation~(\ref{eq:mdotgap_mdotglobal})).
The differences between theirs and our model are within the factor 2.
Since the calculation time of their hydrodynamical simulations is less
than 1/10 of the characteristic viscous evolution time, their values
tends to be larger that those in the steady states.  Thus we expect that
the difference becomes smaller if the calculation time would be longer.
Further investigation by long-term hydrodynamical simulations is
necessary for checking our model.

This gas depletion would also create an inner hole, which is a depleted
region inside a certain radius of a disk \citep[e.g.,][]{Williams11}.
Here we consider a possibility that the inner holes are formed by
planets.
We simply assume in Equation~(\ref{eq:fdash}) that all the gas
approaching to the planet orbit is captured by the planet when
$\dot{M}_{\rm p,hydro} > \dot{M}_{\rm d,global}$ (i.e., in Phase 3), but
Equation~(\ref{eq:fdash2}) means that all the gas is not necessarily
captured even in such a case.
This can be interpreted as the following.  Gas capture by a planet
reduces surface density in wide region, which reduces the gas capture
rate in turn.  When the reduced gas accretion rate is smaller than that
of the global disk-viscous accretion, a fraction of gas that is not
captured by the planet would need to pass through the planet orbit in a
steady state.  This inward flow creates a inner disk with a lower
surface density, which would be a possible origin of the observed inner
holes.  The surface density at the inner hole would be given by $f'
\Sigma_{\rm un}(r)$, whereas we neglected this small amount of mass loss
through the inner hole in phase 3 in this paper.
Note that the gas depletion considered here is different from that by
gap formation, which is created by gravitational torque by the planet
and is usually much narrower.  The gas depletion considered here is a
depletion in addition to that of the gap formation.
Furthermore, the gas depletion due to the gas capture also affects the
type II migration of the planet.  This will be discussed in detail
later.

\begin{figure}
\epsscale{1.00}
\plotone{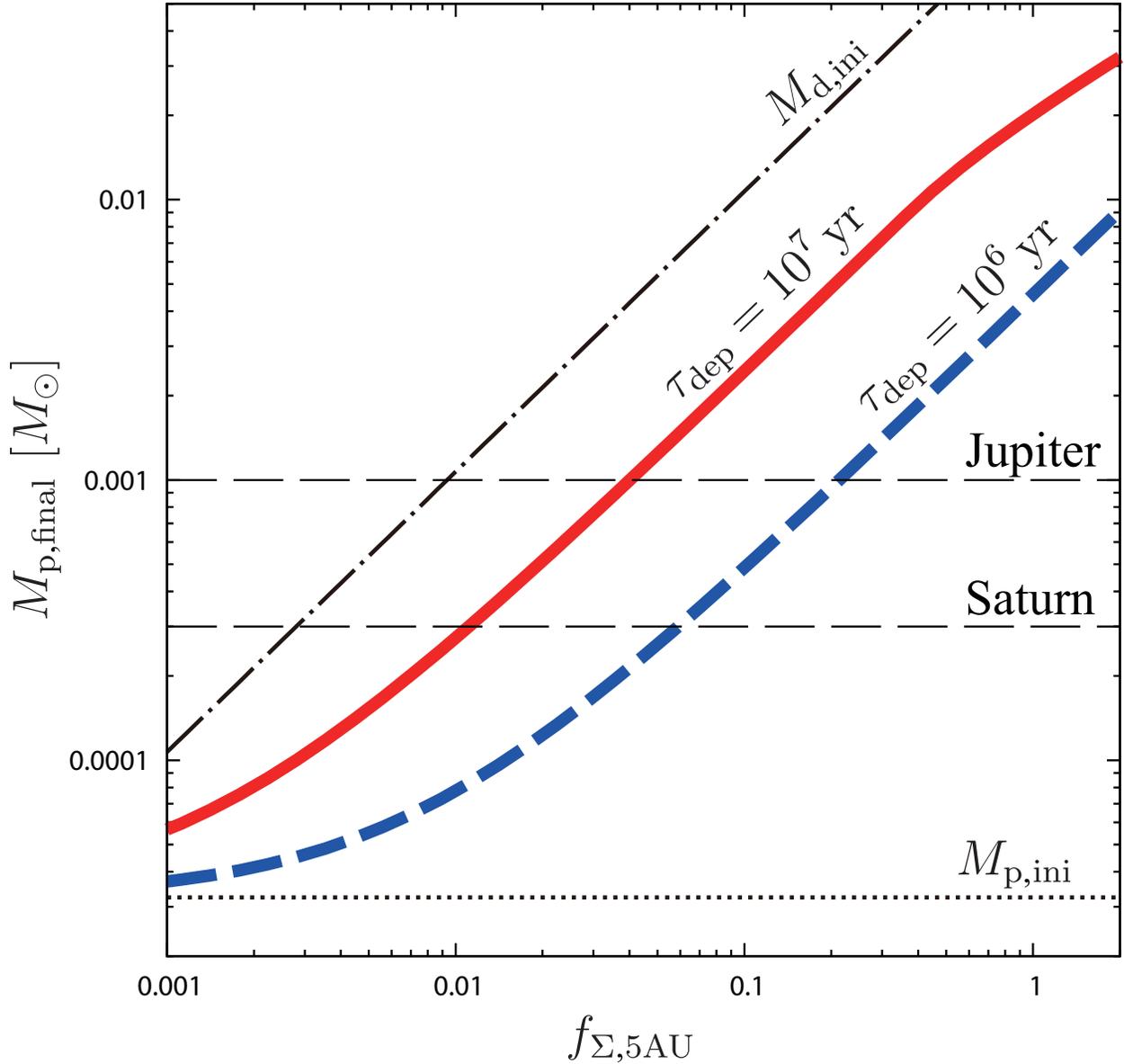}
\caption{Final masses of giant planets at 5AU as a function of
 $f_{\Sigma,\rm 5AU}$ when $\alpha=10^{-3}$, $R_{\rm out} = 200$ AU in
 the cases of $\tau_{\rm dep}=10^7$yr (red solid) and $10^6$yr (blue
 dashed), respectively.  Since $\alpha$ and $\tau_{\rm dep}$ is
 degenerated in most cases, the case with $\tau_{\rm dep} = 10^7$
 corresponds to the case with $\tau_{\rm dep} = 10^6$ and $\alpha =
 10^{-2}$, for example.  The dot-dashed line shows $M_{\rm d,ini}$,
 which corresponds to possible maximum mass of the planet and the dotted
 line shows $M_{\rm p,ini}$.
\label{fig:M_final_vs_f}}
\end{figure}

\clearpage
\section{Implication to the origin of our solar system}\label{sec:implication}
\subsection{A suitable gas disk for Jupiter formation}
In this paper, we updated the model for the growth of giant planets, by
employing the shallow gap model revealed by the recent hydrodynamical
simulations. The updated model showed that the formation of Jupiter-mass
planets requires much less massive gas disks compared to the MMSN model
at the stage of dynamical gas capturing by the planets. This is because
the gap is not so deep to terminate the gas accretion to the planet.

Figure \ref{fig:M_final_vs_f} plots $M_{\rm p,final}$ at 5AU as a
function of $f_{\Sigma,\rm 5AU}$, which indicates the depletion degree
of the disk gas from the MMSN model at the beginning of the gas capture.
The final mass is proportional to the depletion degree, $f_{\Sigma, \rm
5AU}$, when the final mass is much larger than the initial core mass and
smaller than $10M_{\rm J}$ (i.e., within phase 3).  In the case of
$\alpha = 10^{-3}$ and $\tau_{\rm dep}=10^7$yr, a Jupiter-mass planet is
formed in a gas disk with $f_{\Sigma,\rm 5AU} = 0.04$.  The total mass
of this gas disk is about $4M_{\rm J}$.  Such a very low-mass gas disk
also has an advantage that type I and II planetary migrations are both
suppressed significantly.  In a less viscous case $\alpha=10^{-4}$ (or
equivalently a short disk lifetime case $\tau_{\rm dep} = 10^6$yr),
$f_{\Sigma,\rm 5AU}=0.3$ is suitable.  However, such a moderate-mass
disk does not slow down the planetary migrations significantly.  Hence
we adopt the former case as a suitable gas disk for Jupiter formation.
If we adopt a higher viscosity $\alpha > 10^{-3}$ or a longer depletion
time ($\tau_{\rm dep} > 10^7$yr), the suitable disk mass further
decreases.  As mentioned in Section 3, the final mass depends on the
product $\alpha \tau_{\rm dep}$.  Thus, the case of $\alpha=3\e{-3}$ and
$\tau_{\rm dep}=3\e{6}$yr also gives a suitable disk with $f_{\Sigma,\rm
5AU}=0.04$ as well as the former case.
For the Saturn case, $f_{\Sigma,\rm 5AU}$ should be an even lighter disk
($\sim 0.01$ when $\tau_{\rm dep}=10^7$ yr and $\alpha = 10^{-3}$), or
later start (shorter duration) of the dynamical gas capture
(ex. $f_{\Sigma,\rm 5AU} \sim 0.06$ when $\tau_{\rm dep}=10^6$yr).

\subsection{Viscous evolution from a compact disk to a low-mass disk
  with high metallicity}
Jupiter and Saturn have solid cores and also contains a considerable
amount of heavy elements in their H/He envelopes.  The two giant planets
are expected to have solid components of 30-60$M_{\rm E}$ in total
(e.g., \citet{Baraffe14}).  This shows that Jupiter and Saturn have much
higher metallicity than the solar composition.  If we also consider
solids in other planets, the formation of our solar system requires
solid materials of $\gtrsim 80M_{\rm E}$ in total, which is consistent
with the amount of solid included within 50AU of the MMSN disk.
However, the above low-mass disk with $4M_{\rm J}$ contains solid
components of less than $20M_{\rm E}$ if it has the solar composition,
in which heavy elements are 1.4\% in mass.  Hence, the low-mass disk
should have very high metallicity.

A low-mass disk with high metallicity can be formed through a viscous
disk evolution from a compact size described by Equation
(\ref{eq:Sigma_ss}).  Consider an initially compact disk with radius of
$\sim 10$AU and with the solar composition.  The disk mass is $\sim
18M_{\rm J}$, it thus contains solid material of $\sim 80M_{\rm E}$.
Because of the compactness, the gas and solid surface densities of this
disk are twice of the MMSN model at 5AU.  Firstly, planetesimals are
formed in-situ and decoupled with the gaseous disk before the disk
evolution.  Secondly, the compact gas disk suffers a viscous accretion
and losses most of gas within a relatively short time.  The gas disk
spreads out to $\sim$ 200AU and reduces its mass to $\sim 4M_{\rm J}$ at
$t=10^7$yr in the case of $\alpha=10^{-3}$ (see Equation
(\ref{eq:tau_ss})).  During the disk evolution, planetary embryos grow,
and finally a sufficient large solid core causes dynamical collapse of
its envelope and starts to capture its surrounding disk gas rapidly.
This scenario explains the suitable low-mass disk with high metallicity
for Jupiter formation in a natural way.  Saturn formation would also be
reasonable if the onset of dynamical gas capture occurs at even later
time.  It can also naturally explain the high metallicity of Jupiter and
Saturn of the factor $\sim$ 10.

\subsection{Type I and type II migration in gas-depleted
  disks}\label{sec:migration}
Here we discuss more in detail type I and type II migration in
gas-depleted disks.
In particular, we show below that the type II migration of Jupiter-size
planets or smaller is inefficient because of the additional gas
depletion due to the rapid gas capture described in Section
\ref{sec:results}.

Because of the problematic rapid type I migration \citep{Tanaka02}, the
studies on planet population synthesis prefer a gas-depleted (or
high-metallicity) disk or a significant reduction of the type I
migration speed \citep{Daisaka06, IdaLin08a, Mordasini12b}.  In the gas
depleted disk adopted in our scenario, the solid surface density is
twice of the MMSN disk whereas the gas surface density is depleted to
4\% at the end of the core-growth stage.  This requires the solid-to-gas
ratio to be 50 times as large as that of the solar composition.  This
enhancement is comparable to that required in the population synthesis
calculations.  Hence our scenario for Jupiter formation is a plausible
and natural path which can overcome the type I migration problem.
\citet{Mordasini12c} also included the viscous disk evolution in their
population synthesis calculations but they fixed the initial disk outer
radius to be 30AU.  Such an intermediate-size disk takes longer time to
deplete the disk gas enough by disk accretion.  More compact initial
disks should also be examined in population synthesis calculations.

Next we examine type II migration in the gas-depleted disk.  In our
model of giant planet formation, we did not include the effect of type
II migration.  It is worthwhile to estimate the timescale of type II
migration for our gas-depleted disk.  Recently \citet{Duffell14} have
derived an empirical formula of migration speed in the classical type II
regime from their hydrodynamical simulations and the revised timescale
of type II migration is given by
\begin{equation}
  \begin{split}
   t_{\rm migII}
   &= 0.14 \frac{2r_{\rm p}^2}{3\nu_{\rm p}}
   \frac{M_{\rm p}}{\Sigma_{\rm out} r_{\rm p}^2} \\
   &= 0.089 \frac{M_{\rm p}}{\nu_{\rm p} \Sigma_{\rm out}},
  \end{split}
  \label{eq:t_migII}
\end{equation}
where $\Sigma_{\rm out}$ is the gas surface density outside of the gap
and corresponds to $\Sigma(r_{\rm p})$ in Appendix B.  Equation
(\ref{eq:t_migII}) corresponds to the ``planet-dominate'' case where
$M_{\rm p} > \Sigma_{\rm out} r_{\rm p}^2$.  This condition is safely
satisfied in the gas-depleted disk we consider.  Equation
(\ref{eq:t_migII}) agrees well with the hydrodynamical simulations by
\citet{Durmann15} within a factor of $\sim 2$ in the ``planet-dominate''
case and is almost consistent with analytically derived formulae by
\citet{Armitage07}, which is used in population synthesis calculations.

As newly pointed out in this paper, for a giant planet smaller than
$\sim 10M_{\rm J}$, the gas surface density outside of the gap
$\Sigma_{\rm out}$ suffers an additional depletion from the unperturbed
disk because of the rapid gas capture by itself.  The additional
depletion factor $f'$ is given by Equation (\ref{eq:fdash}).  It gives
0.2 for Jupiter mass and $h_{\rm p}/r_{\rm p}=0.05$.  This additional
depletion factor is derived from the mass-flux balance in the outer disk
(see the subsection \ref{sec:evolution} and Appendix B).  This effect
slows down the type II migration and lengthen $t_{\rm migII}$ by the
factor of $1/f'$.  Also note that the additional depletion decreases the
lower limit of $M_{\rm p}$ for the ``planet-dominate'' case.  Any
previous models of type II migration do not include the effect of this
additional gas depletion.  In fact, the gas capture rate, $\dot{M}_{\rm
p,hydro}$, is not taken into account in population synthesis
calculations and thus they cannot evaluate this depletion factor $f' =
\dot{M}_{\rm d,global}/\dot{M}_{\rm p,hydro}$.

\begin{figure}
\epsscale{1.00}
\plotone{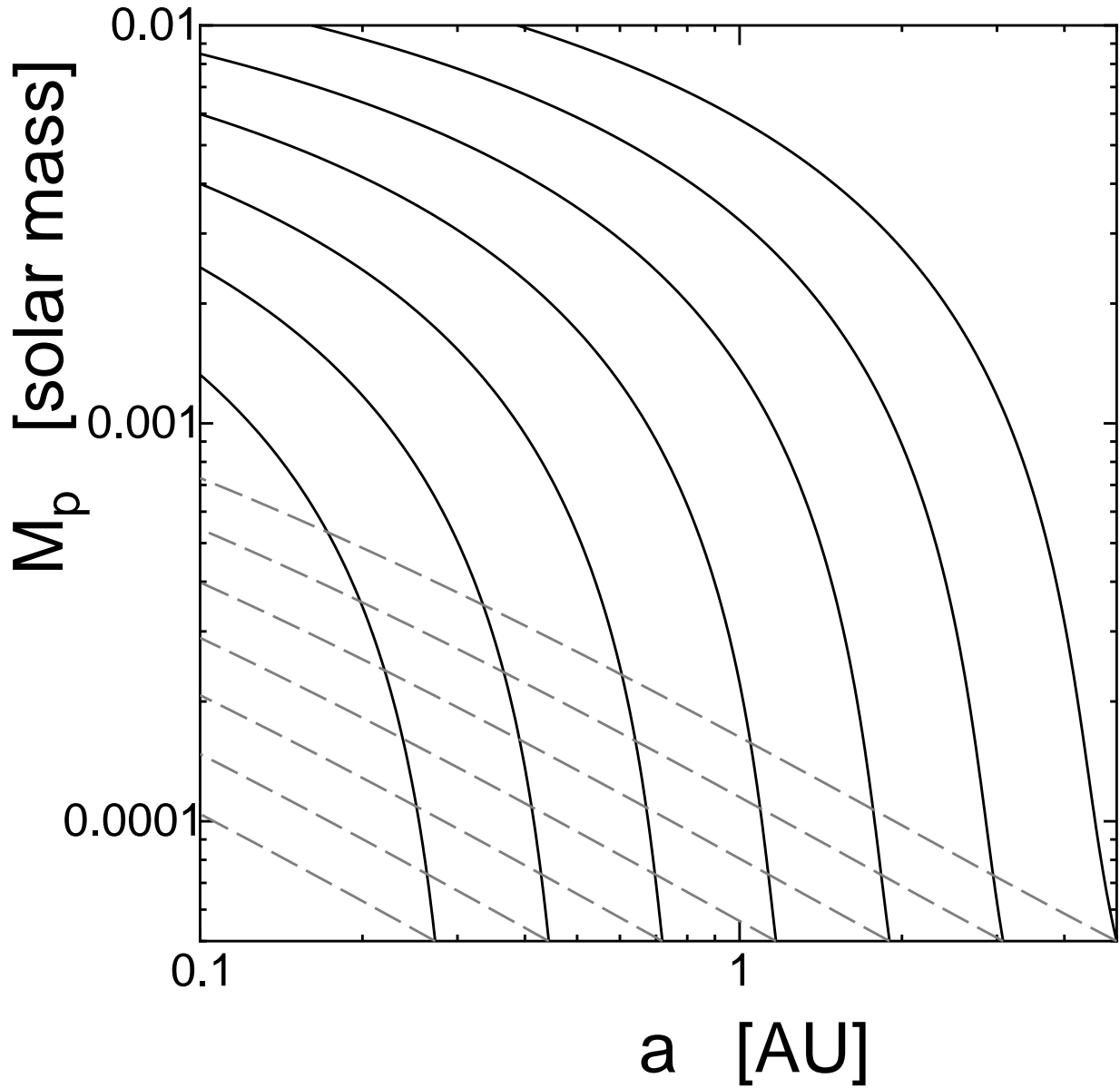}
\caption{Growth-migration curves.  Solid curves show evolution paths
 base on our result (Equation~(\ref{eq:tgrow_per_tmigII})), and dashed
 curves are the cases where the traditional type II migration is used.
\label{fig:M_p_vs_a}}
\end{figure}

%
The growth time of giant planets less than $\sim 10M_{\rm J}$ is given
by
\begin{equation}
 t_{\rm grow}
  = \frac{M_{\rm p}}{\dot{M}_{\rm p}}
  = \frac{M_{\rm p}}{3\pi \nu_{\rm p} \Sigma_{\rm un}},
 \label{eq:tgrow}
\end{equation}
which is almost equal to $t_{\rm migII}$ if the unperturbed surface
density $\Sigma_{\rm un}$ is replaced by $\Sigma_{\rm out}$.  Hence the
ratio $t_{\rm grow}/t_{\rm migII}$ is given by
\begin{equation}
 \frac{t_{\rm grow}}{t_{\rm migII}}
  = f'
  \simeq 1.2 \times \kakkoi{M_{\rm p}}{M_\ast}{2/3}
                    \kakkoi{h_{\rm p}}{r_{\rm p}}{-1}.
 \label{eq:tgrow_per_tmigII}
\end{equation}
This indicates the Jupiter mass or smaller planets suffer only a small
radial drift by type II migration during their growth.  According to our
growth model, the growth of giant planets terminates only when the disk
is depleted to a negligible mass.  Hence our results indicate that type
II migration is ineffective for Jupiter mass planets or smaller.  On the
other hand, since the previous models neglect this additional gas
depletion, the growth time is always comparable to $t_{\rm migII}$,
provided that the growth is controlled by the global disk accretion
\citep{Benz14}.
Figure~\ref{fig:M_p_vs_a} shows growth-migration curves.  Evolutions of
our model shows growth without significant migration until $M_{\rm p}
\sim M_{\rm J}$, while the case with traditional type II migration shows
relatively rapid inward migration.

Calculations of planet population synthesis suggest that most giant
planets fall to their host stars because of their rapid type II
migration \citep[e.g.,][]{IdaLin08b}. This is inconsistent with the
radial distribution of observed extra-solar planets, in which
Jupiter-sized planets are piled up at $\sim$1AU and hot Jupiters are
minor \citep{Mayor11, Ida13}.  \citet{Hasegawa13} discussed possible
mechanisms which slows down the type II migration but did not find any
effective slowdown mechanisms. In the above and
Figure~\ref{fig:M_p_vs_a}, we showed that type II migration of giant
planets smaller than $\sim 10M_{\rm J}$ slows down because of the
additional gas depletion due to their rapid gas capture. Our slow down
mechanism may resolve the problem of type II migration.

For giant planets larger than $\sim 10M_{\rm J}$ (i.e., in phase 4), the
gap effect prolongs the growth time compared with Equation
(\ref{eq:tgrow}) and the additional gas depletion does not occur.  Then
we find that the time ratio is again given by Equation
(\ref{eq:tgrow_per_tmigII}) but $f'$ given by Equation (\ref{eq:fdash})
is larger than unity in this case.  The time scale of type II migration
is shorter than the growth time for $M_{\rm p} > 10M_{\rm J}$. This may
explain that extra-solar planets more massive than $10M_{\rm J}$ are
observed less frequently.  A detail population synthesis calculation
would be necessary, including our slowdown mechanism for type II
migration.

\section{Summary and Discussion}\label{sec:summary}
We examined the growth rates and the final masses of giant planets
embedded in protoplanetary disks through capture of disk gas, by
employing an empirical formula for the gas capture rate and a shallow
disk gap model, which are both based on hydrodynamical simulations. Our
findings are summarized as follows.

\begin{enumerate}
 \item Because of the shallow gap revealed by recent hydrodynamical
       simulations, giant planets do not stop their gas-capturing
       growth.  Only the depletion of the whole gas disk can terminate
       their growth.  For planets less massive than $10 M_{\rm J}$, their
       growth rates are mainly controlled by the gas supply through the
       global disk accretion, rather than their gaps.  For such a mass
       range, the final planet mass is given by Equation
       (\ref{eq:M_p_final_p3}). For the more massive planets, their
       growth rates are limited by deep gaps and their final mass is
       given by Equation (\ref{eq:M_p_final_p4}).

 \item For planets less massive than $10 M_{\rm J}$, the gas supply to the
       planets by the disk accretion is insufficient. This also causes a
       depletion of the gas surface density even at the outside of the
       gap and create an inner hole in the protoplanetary disk The
       additional gas depletion factor is given by Equation
       (\ref{eq:fdash}) (see also Appendix B).  Our result suggests that
       less massive giant planets can create deeper inner holes than
       more massive ones.

 \item Because of the non-stopping growth, the Jupiter formation
       requires a very low-mass gas disk with a few or several $M_{\rm
       J}$, at least, at the beginning of its gas capture.  This disk is
       much less massive than the MMSN model, whereas the solid material
       of $\sim$ 80 Earth masses is also necessary for formation of the
       planets in our solar system.  That is, we need a very low-mass
       disk with a high metallicity. These requirements can be achieved
       by the viscous evolution from an initially $\sim$ 10AU-sized
       compact disk with the solar composition. For a disk with a
       moderate viscosity of $\alpha \sim 10^{-3}$, most of disk gas
       accretes onto the central star and a widely-spread low-mass gas
       disk remains at $t\sim 10^7$ yrs.  This scenario can explain the
       high metallicity in giant planets of our solar system.

 \item A very low-mass gas disk also provides a plausible path where
       type I and II planetary migrations are both suppressed
       significantly.  In particular, we also showed that the type II
       migration of Jupiter-size planets is inefficient because of the
       additional gas depletion due to their rapid gas capture.
       This slow type II migration is consistent with the radial
       distribution of observed extra-solar planets in which
       Jupiter-sized planets are piled up at $\sim 1$AU.
\end{enumerate}

%
In this paper, we proposed the formula that describes the gas accretion
rate onto the planet, which combines an empirical formula of gas
accretion rate obtained by a local hydrodynamic simulation (TW02) and
another empirical formula of gap depth obtained by global hydrodynamical
simulation \citep{DM13}.
Although the accretion rate is in good agreement with global
hydrodynamic simulations, the function $\dot{M}_{\rm p,hydro}$ (or more
specifically $D$) should be checked by making use of recent advancement
of hydrodynamic simulations.
We also proposed a formation of large-scale disk depletion created by
gas capture of the planet based on analytic argument (see appendix B).
This depletion arises also from long-term viscous evolution of the gas
disk, which have not been studies so far.
To check this, high-resolution hydrodynamic simulations with long-term
evolution should be done in future work.
\citet{Malik15} have recently investigated the gap-opening criterion of
migrating planets in protoplanetary disks and found that the gap opening
is more difficult than the traditional condition, which is based on the
torque balance \citep[e.g.,][]{Crida06, LP93}.
Other recent hydrodynamic simulations \citep{DM13, Fung14} also showed
that the gap opening is less significant, although there are some
differences on assumptions and purposes.
Also, gas in the gap region tends to be turbulence by magneto-rotational
instability \citep{Gressel13, Keith15}, which would lead to an even
shallower gap.
To quantify the gap deepening effect, further studies would be
necessary.

In-situ formation of hot jupiters is thought to be difficult in general.
This is because (1) a planet in inner region is easier to form a gap
than that in outer region, which meant that the planet growth stops at a
mass much less than that of Jupiter, (2) even if a hot jupiter forms, it
is susceptible to type II migration and falls to the central star, and
(3) there are not enough solid materials to trigger the dynamical gas
capture of the protoplanet.
Our model showed that the gap has little effect on suppressing gas
accretion rate onto planets.  In this sense, the first point is not a
reason to prevent from in-situ formation anymore.  We also showed that
the type II migration is less effective than previously thought, which
would not rule out the in-situ formation either.
Although we have no claim that there are enough solid to form large
solid cores, we should be noted that the difficulty of the in-situ
formation of hot jupiters are greatly mitigated.

Our model assumes forced exponential decay with respect to time for the
protoplanetary disk, but the way of disk dissipation would impact
significantly on disk evolution and thus also on the planet formation
scenario described above.
Several mechanisms in addition to viscous disk accretion are proposed,
such as photoevaporation \citep{Hollenbach94, Clarke01, Owen12}, solar
wind stripping \citep{Horedt78, Matsuyama09}, solar wind induced
accretion to the central star \citep{Elmegreen78}, disk wind
\citep{Suzuki09, Suzuki14}.
Among them, photoevaporation is thought to be the dominant mechanism to
dissipate disks.

%
%
Photoevaporation has been actively studied recently and roles of X-ray
and UV irradiation from the central star has been substantially
understood.
Since, however, luminosity of X-ray, EUV, and FUV and the time evolution
has large uncertainty, the evolution of protoplanetary disks would have
wide variety.
If disk accretion rate toward the central star:
\begin{equation}
\dot{M}_{\rm d,global}
 = 5\e{-9}
   f_{\Sigma,\rm 5AU}
   \kakkoi{\alpha}{10^{-3}}{1}
   \kakkoi{h_{\rm 1AU}/{\rm 1AU}}{10^{-1.5}}{2}
   M_\odot / {\rm yr}
\end{equation}
is larger than mass-loss rate by photoevaporation, the effect of
photoevaporation can be neglected.
Our scenario suggests that $f_{\Sigma,\rm 5AU} \sim 0.04$ is a plausible
parameter, which gives $\dot{M}_{\rm d,global} = 2\e{-10} M_\odot/$yr.
According to \citet{Owen12}, mass-loss rate can be in a range from
$10^{-12}$ to $10^{-7} M_\odot$/yr depending on X-ray luminosity, which
means that our scenario is realized when X-ray luminosity is not strong.


%


We are grateful to Masahiro Ikoma, Hiroshi Kobayashi, Satoshi Okuzumi,
and Kazuhiro Kanagawa for valuable comments.
This work was supported by JSPS KAKENHI Grant Numbers 24103503, 26800229,
15H02065, 26287101.
%
%

\clearpage
\appendix
\section{Derivation of Equation (\ref{eq:mdotdl})}
The diffusion equation of surface density $\Sigma(r,t)$ in a Keplerian
disk with viscosity $\nu$ is
\begin{equation}
 \pfrac{\Sigma}{t}
  = \frac{3}{r}
  \pfrac{}{r}
  \left[ r^{1/2}
   \pfrac{}{r}
   (\Sigma \nu r^{1/2})
  \right].
\end{equation}
Adopting the local approximation ($x=r-r_{\rm p} \ll 1$), we can rewrite
the diffusion equation as
\begin{equation}
 \pfrac{\Sigma}{t}
  = 3 \nu_{\rm p} \ppfrac{\Sigma}{x}.
\end{equation}
Under the following initial and boundary conditions:
\begin{gather}
 \Sigma(x,t=0) = \Sigma_{\rm un}, \\
 \Sigma(x=0,t) = 0,
\end{gather}
we obtain the exact solution \citep[e.g.,][]{Landau_Fluid}:
\begin{equation}
 \Sigma(x,t)
  = \Sigma_{\rm un} \, \erf \left( \frac{x}{2\sqrt{3\nu_{\rm p} t}}\right),
\end{equation}
where $\Sigma_{\rm un}$ is a constant.

Since the disk mass reduced from the initial condition equals to the
mass sunk into the boundary at $x=0$, we have
\begin{align}
 \int_0^t F(t') dt'
 &= 2\pi r_{\rm p} \int_0^\infty (\Sigma_{\rm un} - \Sigma(x,t))) dx \nonumber \\
 &= 2\pi r_{\rm p} \Sigma_{\rm un} \frac{2\sqrt{3\nu_{\rm p}
 t}}{\sqrt{\pi}},
 \label{eq:F_integ}
\end{align}
where $F(t)$ is mass flux toward $x=0$ from $x>0$, and we use a
mathematical formula: $\int_0^\infty \erfc(x) dx = 1/\sqrt{\pi}$.
Differentiating Equation (\ref{eq:F_integ}) with respect to time yields
the mass accretion rate to the boundary $x=0$ as a function of time:
\begin{align}
 F(t)
 &= 2\pi r_{\rm p} \Sigma_{\rm un} \sqrt{\frac{3\nu_{\rm p}}{\pi t}}.
\end{align}

\clearpage
\section{A solution of a steady accretion disk with a mass sink to an embedded planet}
\citet{Lubow06} obtained the radial gas distribution of a steady viscous
accretion disk with a mass sink by a planet
and showed that the mass sink causes a gas depletion in a wide region,
but did not give a detailed derivation of it.  Here, we present the
derivation the solution of the surface density distribution in such a
case.
In an accretion disk, the radial angular moment flux $F_J$ is given by
\citep{LP74}
\begin{equation}
F_J = j F_M + 3\pi r^2 \nu \Sigma \Omega,
\label{fj}
\end{equation}
where $F_M$ is the radial mass flux and the specific angular momentum
$j$ is given by $r^2 \Omega$. For $\Omega$, the Keplerian rotation is
assumed.  Thus, for given (constant) mass flux and angular momentum
flux, the surface density of the quasi-steady disk is expressed as
\begin{equation}
\Sigma(r) = { -F_M + F_J/j(r) \over 3 \pi \nu(r) }.
\label{sigma}
\end{equation}

We here consider an accretion disk having a mass sink with the rate of
$\dot{M}_{\rm p}$ at $r=r_{\rm p}$ due of the accretion onto the planet.  The disk
angular momentum sinks into the planet with the rate of $j(r_{\rm p})
\dot{M}_{\rm p}$.  Then the mass and angular momentum fluxes are discontinuous
at $r_{\rm p}$ and given by
\begin{equation}
F_M = 
\left\{ 
\begin{array}{lc}
-(\dot{M}_\ast+\dot{M}_{\rm p}) & (r>r_{\rm p}),\\[2mm]
-\dot{M}_\ast & (r<r_{\rm p}),\\
\end{array}
\right .
\label{fm1}
\end{equation}
and
\begin{equation}
F_J = 
\left\{ 
\begin{array}{lc}
-(j_\ast \dot{M}_\ast + j_{\rm p} \dot{M}_{\rm p}) & (r>r_{\rm p}),\\[2mm]
-j_\ast\dot{M}_\ast & (r<r_{\rm p}),\\
\end{array}
\right .
\label{fj1}
\end{equation}
where $\dot{M}_\ast$ is the mass accretion rate onto the central star and
$r_\ast$ is the radius of the inner disk edge. The specific angular
momenta $j_{\rm p}$ and $j_\ast$ are the values at $r_{\rm p}$ and $r_\ast$,
respectively.  Note that the sum $\dot{M}_{\rm p}+ \dot{M}_\ast$ is equal to the
global accretion rate $\dot{M}_{\rm d,global}$. The negative fluxes
indicate the inward transport of the disk mass and angular
momentum. Substituting Equations (\ref{fm1}) and (\ref{fj1}) into
(\ref{sigma}), we obtain
\begin{equation}
\Sigma(r) = 
\left\{ 
\begin{array}{lc}
\displaystyle
{ \dot{M}_\ast \over 3 \pi \nu } \left(1-\sqrt{r_\ast \over r}\right) 
+ { \dot{M}_{\rm p} \over 3 \pi \nu} \left(1-\sqrt{r_{\rm p} \over r}\right) 
&\mbox{for}\quad r>r_{\rm p},\\[4mm]
\displaystyle
{\dot{M}_\ast \over 3 \pi \nu} 
             \left(1-\sqrt{r_\ast \over r}\right)
&\mbox{for}\quad r<r_{\rm p}.\\
\end{array}
\right .
\label{sigma2}
\end{equation}

\begin{figure}
\epsscale{1.00}
\plotone{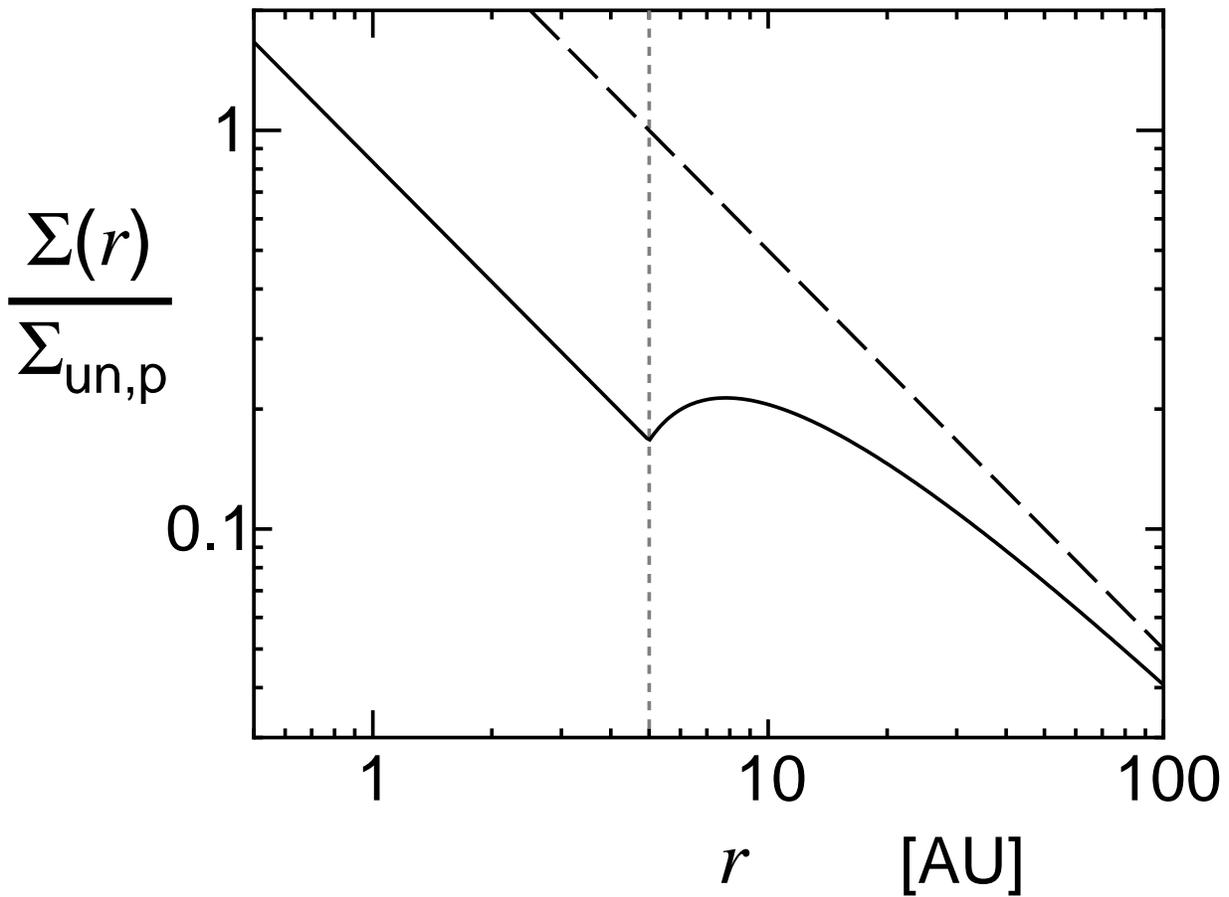}
\caption{An example of the surface density distribution of the disk with 
a mass sink of the planetary gas capture (the solid line).
The planet is located at 5AU (the vertical gray dotted line).
The ratio $ D/3\pi \nu_{\rm p}$ is set to be 5. 
It is also assumed that $r \gg r_\ast$.
The unperturbed surface density
$\Sigma_{un}$ is also plotted by the dashed line for comparison.
Because of the planetary gas capture, the gas surface density is further
depleted and an inner hole is formed. 
\label{fig:sd}}
\end{figure}

The ratio $\dot{M}_\ast / \dot{M}_{\rm p}$ is determined by the accretion
formula of Equation~(\ref{eq:mdotphydro}).  From
Equation~(\ref{sigma2}), we obtain
\begin{equation}
\Sigma(r_{\rm p}) = {\dot{M}_\ast \over 3\pi \nu_{\rm p}},
\label{sigmap}
\end{equation}
where we omitted the term of $\sqrt{r_\ast/r_{\rm p}}$.  This approximation
would be valid for a planet with $r_{\rm p} \gtrsim 1$AU since $r_\ast$ would
be less than 0.1AU.  Note that the solution given by \citet{Lubow06}
does not assume that $r_\ast \ll r_{\rm p}$.  From Equations
(\ref{eq:mdotphydro}), (\ref{eq:Sigma_acc_DM13}) and (\ref{sigmap}), we
obtain
\begin{equation}
\dot{M}_{\rm p} = {D' \over 3\pi \nu_{\rm p}} \dot{M}_\ast,
\label{mdotp0}
\end{equation}
where $D'$ is defined by
\begin{equation}
D' = { 1 \over 0.034 K +1 } D,
\label{Ddashdef}
\end{equation}
$\nu_{\rm p} = \nu(r_{\rm p})$, and we equate $\Sigma(r_{\rm p})$ with
$\Sigma_{\rm un}$ in Equation~(\ref{eq:Sigma_acc_DM13}).  Since the
ratio $D' / 3\pi \nu_{\rm p}$ is equal to $\dot{M}_{\rm
p,hydro}/\dot{M}_{\rm d,global}$ by definition, the ratio is given by
Equation (\ref{eq:mdotgap_mdotglobal}):
\begin{equation}
{D' \over 3\pi \nu_{\rm p}} = 
0.90 \left( {M_{\rm p} \over M_*} \right)^{-2/3}  \left( {h_{\rm p}
      \over r_{\rm p}} \right)
= 4.5 \left( {M_{\rm p} \over M_J} \right)^{-2/3} \left( {h_{\rm p}
       /r_{\rm p} \over 0.05} \right).
\end{equation}
Noting $\dot{M}_{\rm p} + \dot{M}_\ast = \dot{M}_{\rm d,global}$ and using 
Equation~(\ref{mdotp0}), we obtain the accretion rates as
\begin{equation}
\dot{M}_{\rm p} = { D'/3\pi \nu_{\rm p} \over 1+ D'/3\pi \nu_{\rm p}}
 \dot{M}_{\rm d,global}, \qquad 
\dot{M}_\ast = { 1 \over 1+ D'/3\pi \nu_{\rm p}} \dot{M}_{\rm d,global}.
\label{mdots}
\end{equation}
Substituting Equation (\ref{mdots}) into (\ref{sigma2}),
we finally obtain the expression of the surface density of the disk with 
a mass sink of the planetary gas capture as
\begin{equation}
\Sigma(r) = 
\left\{ 
\begin{array}{lc}
\displaystyle
{ \dot{M}_{\rm d,global} \over 3 \pi \nu (1+ D'/3\pi \nu_{\rm p})}
             \left[  
\left(1-\sqrt{r_\ast \over r}\right) 
+
{ D' \over 3\pi \nu_{\rm p}}\left(1-\sqrt{r_{\rm p} \over r}\right) 
\right]
&\mbox{for}\quad r>r_{\rm p},\\[6mm]
\displaystyle
{\dot{M}_{\rm d,global} \over 3 \pi \nu (1+ D'/3\pi \nu_{\rm p})} 
             \left(1-\sqrt{r_\ast \over r}\right)
&\mbox{for}\quad r<r_{\rm p}.\\
\end{array}
\right .
\label{sigma3}
\end{equation}
From the expression, the additional gas depletion factor due to the gas
capture is given by $(1+ D'/3\pi \nu_{\rm p})^{-1}$, which is
approximately equal to Equation (\ref{eq:fdash2}) in the text.  In
Figure~\ref{fig:sd} we plot the obtained surface density distribution
for a typical case.
In the vicinity of the planet, the gap, which is an additional gas
depletion due to the gravitational torque from the planet, should also
exist.
Even though the obtained surface density does not show this gap, the
effect of the gap is included in this formulation through the parameter
$D'$ of Equation (\ref{Ddashdef}).

\clearpage

\end{document}